\documentstyle[11pt,qqaaelba,epsf]{article}

\newcommand{\cf}{{\it cf.\ }}

\newcommand{\alfaS}{\mbox{$\alpha_{s}$}}

\newcommand{\Msbar}{\mbox{\small {$\overline {\rm MS}$}}}

\catcode`\@=11

\let\DOTSI\relax
\def\RIfM@{\relax\ifmmode}
\def\FN@{\futurelet\next}
\newcount\intno@
\def\iint{\DOTSI\intno@\tw@\FN@\ints@}
\def\iiint{\DOTSI\intno@\thr@@\FN@\ints@}
\def\iiiint{\DOTSI\intno@4 \FN@\ints@}
\def\idotsint{\DOTSI\intno@\z@\FN@\ints@}
\def\ints@{\findlimits@\ints@@}
\newif\iflimtoken@
\newif\iflimits@
\def\findlimits@{\limtoken@true\ifx\next\limits\limits@true
 \else\ifx\next\nolimits\limits@false\else
 \limtoken@false\ifx\ilimits@\nolimits\limits@false\else
 \ifinner\limits@false\else\limits@true\fi\fi\fi\fi}
\def\multint@{\int\ifnum\intno@=\z@\intdots@                                
 \else\intkern@\fi                                                          
 \ifnum\intno@>\tw@\int\intkern@\fi                                         
 \ifnum\intno@>\thr@@\int\intkern@\fi                                       
 \int}                                                                      
\def\multintlimits@{\intop\ifnum\intno@=\z@\intdots@\else\intkern@\fi
 \ifnum\intno@>\tw@\intop\intkern@\fi
 \ifnum\intno@>\thr@@\intop\intkern@\fi\intop}
\def\intic@{\mathchoice{\hskip.5em}{\hskip.4em}{\hskip.4em}{\hskip.4em}}
\def\negintic@{\mathchoice
 {\hskip-.5em}{\hskip-.4em}{\hskip-.4em}{\hskip-.4em}}
\def\ints@@{\iflimtoken@                                                    
 \def\ints@@@{\iflimits@\negintic@\mathop{\intic@\multintlimits@}\limits    
  \else\multint@\nolimits\fi                                                
  \eat@}                                                                    
 \else                                                                      
 \def\ints@@@{\iflimits@\negintic@
  \mathop{\intic@\multintlimits@}\limits\else
  \multint@\nolimits\fi}\fi\ints@@@}
\def\intkern@{\mathchoice{\!\!\!}{\!\!}{\!\!}{\!\!}}
\def\plaincdots@{\mathinner{\cdotp\cdotp\cdotp}}
\def\intdots@{\mathchoice{\plaincdots@}
 {{\cdotp}\mkern1.5mu{\cdotp}\mkern1.5mu{\cdotp}}
 {{\cdotp}\mkern1mu{\cdotp}\mkern1mu{\cdotp}}
 {{\cdotp}\mkern1mu{\cdotp}\mkern1mu{\cdotp}}}

\newif\iffirstchoice@
\firstchoice@true
\def\textfonti{\the\textfont\@ne}
\def\textfontii{\the\textfont\tw@}
\def\text{\RIfM@\expandafter\text@\else\expandafter\text@@\fi}
\def\text@@#1{\leavevmode\hbox{#1}}
\def\text@#1{\mathchoice
 {\hbox{\everymath{\displaystyle}\def\textfonti{\the\textfont\@ne}%
  \def\textfontii{\the\textfont\tw@}\textdef@@ T#1}}
 {\hbox{\firstchoice@false
  \everymath{\textstyle}\def\textfonti{\the\textfont\@ne}%
  \def\textfontii{\the\textfont\tw@}\textdef@@ T#1}}
 {\hbox{\firstchoice@false
  \everymath{\scriptstyle}\def\textfonti{\the\scriptfont\@ne}%
  \def\textfontii{\the\scriptfont\tw@}\textdef@@ S\rm#1}}
 {\hbox{\firstchoice@false
  \everymath{\scriptscriptstyle}\def\textfonti
  {\the\scriptscriptfont\@ne}%
  \def\textfontii{\the\scriptscriptfont\tw@}\textdef@@ s\rm#1}}}
\def\textdef@@#1{\textdef@#1\rm\textdef@#1\bf\textdef@#1\sl\textdef@#1\it}
\def\DN@{\def\next@}
\def\eat@#1{}
\def\textdef@#1#2{%
 \DN@{\csname\expandafter\eat@\string#2fam\endcsname}%
 \if S#1\edef#2{\the\scriptfont\next@\relax}%
 \else\if s#1\edef#2{\the\scriptscriptfont\next@\relax}%
 \else\edef#2{\the\textfont\next@\relax}\fi\fi}

\def\Let@{\relax\iffalse{\fi\let\\=\cr\iffalse}\fi}
\def\vspace@{\def\vspace##1{\crcr\noalign{\vskip##1\relax}}}
\def\multilimits@{\bgroup\vspace@\Let@
 \baselineskip\fontdimen10 \scriptfont\tw@
 \advance\baselineskip\fontdimen12 \scriptfont\tw@
 \lineskip\thr@@\fontdimen8 \scriptfont\thr@@
 \lineskiplimit\lineskip
 \vbox\bgroup\ialign\bgroup\hfil$\m@th\scriptstyle{##}$\hfil\crcr}
\def\Sb{_\multilimits@}
\def\endSb{\crcr\egroup\egroup\egroup}
\def\Sp{^\multilimits@}

\newdimen\ex@
\def\rightarrowfill@#1{$#1\m@th\mathord-\mkern-6mu\cleaders
 \hbox{$#1\mkern-2mu\mathord-\mkern-2mu$}\hfill
 \mkern-6mu\mathord\rightarrow$}
\def\leftarrowfill@#1{$#1\m@th\mathord\leftarrow\mkern-6mu\cleaders
 \hbox{$#1\mkern-2mu\mathord-\mkern-2mu$}\hfill\mkern-6mu\mathord-$}
\def\leftrightarrowfill@#1{$#1\m@th\mathord\leftarrow\mkern-6mu\cleaders
 \hbox{$#1\mkern-2mu\mathord-\mkern-2mu$}\hfill
 \mkern-6mu\mathord\rightarrow$}
\def\overrightarrow{\mathpalette\overrightarrow@}
\def\overrightarrow@#1#2{\vbox{\ialign{##\crcr\rightarrowfill@#1\crcr
 \noalign{\kern-\ex@\nointerlineskip}$\m@th\hfil#1#2\hfil$\crcr}}}

\def\overleftarrow{\mathpalette\overleftarrow@}
\def\overleftarrow@#1#2{\vbox{\ialign{##\crcr\leftarrowfill@#1\crcr
 \noalign{\kern-\ex@\nointerlineskip}$\m@th\hfil#1#2\hfil$\crcr}}}
\def\overleftrightarrow{\mathpalette\overleftrightarrow@}
\def\overleftrightarrow@#1#2{\vbox{\ialign{##\crcr\leftrightarrowfill@#1\crcr
 \noalign{\kern-\ex@\nointerlineskip}$\m@th\hfil#1#2\hfil$\crcr}}}
\def\underrightarrow{\mathpalette\underrightarrow@}
\def\underrightarrow@#1#2{\vtop{\ialign{##\crcr$\m@th\hfil#1#2\hfil$\crcr
 \noalign{\nointerlineskip}\rightarrowfill@#1\crcr}}}

\def\underleftarrow{\mathpalette\underleftarrow@}
\def\underleftarrow@#1#2{\vtop{\ialign{##\crcr$\m@th\hfil#1#2\hfil$\crcr
 \noalign{\nointerlineskip}\leftarrowfill@#1\crcr}}}
\def\underleftrightarrow{\mathpalette\underleftrightarrow@}
\def\underleftrightarrow@#1#2{\vtop{\ialign{##\crcr$\m@th\hfil#1#2\hfil$\crcr
 \noalign{\nointerlineskip}\leftrightarrowfill@#1\crcr}}}

\catcode`\@=\active

\def\frac#1#2{{#1 \over #2}}

\def\dfrac#1#2{{\displaystyle {#1 \over #2}}}

\input{epsf.tex}
\def\FiguresIn#1{\def\@Figures{#1}}

\newcount\GRAPHICSTYPE
\def\graffile#1#2#3#4{\leavevmode\raise -#4 \hbox{%
\raise #3 \hbox{\rule{0.003in}{0.003in}\special{#1}}}%
{\raise -#4 \hbox to #2 {\vrule height#3 width0in depth0in\hfil}}%
}

\def\draftbox#1#2#3#4{\leavevmode\raise -#4 \hbox{\frame{\rlap{\protect\tiny #1}%
\hbox to #2{\vrule height#3 width0in depth0in\hfil}}}}

\newcount\draft
\def\GRAPHIC#1#2#3#4#5{\ifnum\draft=1 \draftbox{#2}{#3}{#4}{#5}\else%
\graffile{#1}{#3}{#4}{#5}\fi}

\def\addtoLaTeXparams#1{\edef\LaTeXparams{\LaTeXparams #1}}

\def\doFRAMEparams#1{\readFRAMEparams#1\end}
\def\readFRAMEparams#1{%
\ifx#1\end%
\let\next=\relax%
\else%
\ifx#1i%
\dispkind=0%
\fi%
\ifx#1d%
\dispkind=1%
\fi%
\ifx#1f%
\dispkind=2%
\fi%
\ifx#1t%
\addtoLaTeXparams{t}%
\fi%
\ifx#1b%
\addtoLaTeXparams{b}%
\fi%
\ifx#1p%
\addtoLaTeXparams{p}%
\fi%
\ifx#1h%
\addtoLaTeXparams{h}%
\fi%
\let\next=\readFRAMEparams%
\fi%
\next%
}

\def\IFRAME#1#2#3#4#5{\GRAPHIC{#5}{#4}{#1}{#2}{#3}}

\def\DFRAME#1#2#3#4{
  \begin{center}
    \GRAPHIC{#4}{#3}{#1}{#2}{0in} 
  \end{center}
}

\def\FFRAME#1#2#3#4#5#6#7{
  \begin{figure}[#1]
    \begin{center}
      \leavevmode\epsffile{\@Figures#7}
    \end{center}
    \caption{\label{#5}#4}
  \end{figure}
}

\def\FRAME#1#2#3#4#5#6#7#8{%
\newcount\dispkind%
\def\LaTeXparams{}%
\dispkind=0%
\def\LaTeXparams{}%
\doFRAMEparams{#1}%
\ifnum\dispkind=0%
\IFRAME{#2}{#3}{#4}{#7}{#8}%
\else
  \ifnum\dispkind=1
    \DFRAME{#2}{#3}{#7}{#8}
  \else
    \ifnum\dispkind=2
      \FFRAME{\LaTeXparams}{#2}{#3}{#5}{#6}{#7}{#8}
    \fi
  \fi
\fi
}

\catcode`\@=11

\long\def\QQQ#1#2{}
\def\QTP#1{}
\long\def\QQA#1#2{}

\def\EXPAND#1[#2]#3{}
\def\NOEXPAND#1[#2]#3{}

\def\LaTeXparent#1{}

\@ifundefined{INDEX}{\def\INDEX#1#2{}{}}{}
\@ifundefined{SUBINDEX}{\def\SUBINDEX#1#2#3{}{}{}}{}
\def\initial#1{\bigbreak{\raggedright\large\bf #1}\kern 2pt\penalty3000}

\@ifundefined{abstract}{%
\def\abstract{\if@twocolumn
\section*{Abstract (Not appropriate in this style!)}
\else \small 
\begin{center}
{\bf Abstract\vspace{-.5em}\vspace{0pt}} 
\end{center}
\quotation 
\fi}}{}
\@ifundefined{endabstract}{%
\def\endabstract{\if@twocolumn\else\endquotation\fi}}{}
\@ifundefined{maketitle}{\def\maketitle#1{}}{}
\@ifundefined{affiliation}{\def\affiliation#1{}}{}
\@ifundefined{proof}{}{}
\@ifundefined{newfield}{\def\newfield#1#2{}}{}
\@ifundefined{chapter}{\def\chapter#1{\par(Chapter head:)#1\par }}{}
\@ifundefined{part}{\def\part#1{\par(Part head:)#1\par }}{}
\@ifundefined{section}{\def\part#1{\par(Section head:)#1\par }}{}
\@ifundefined{subsection}{\def\part#1{\par(Subsection head:)#1\par }}{}
\@ifundefined{subsubsection}{\def\part#1{\par(Subsubsection head:)#1\par }}{}
\@ifundefined{paragraph}{\def\part#1{\par(Subsubsubsection head:)#1\par }}{}
\@ifundefined{subparagraph}{\def\part#1{\par(Subsubsubsubsection head:)#1\par }}{}

\newdimen\theight
\def \Column{%
             \vadjust{\setbox0=\hbox{\scriptsize\quad\quad tcol}%
             \theight=\ht0
             \advance\theight by \dp0    \advance\theight by \lineskip
             \kern -\theight \vbox to \theight{\rightline{\rlap{\box0}}%
             \vss}%
             }}%

\def\qed{\ifhmode\unskip\nobreak\fi\ifmmode\ifinner\else\hskip5\p@\fi\fi
 \hbox{\hskip5\p@\vrule width4\p@ height6\p@ depth1.5\p@\hskip\p@}}
\catcode`@=12 

\makeatletter

\def\cf{{\it cf.}}

\begin{document}

\def\macgraphsize#1#2#3{\vbox to #2{\hsize #1\relax%
\hrule height 0pt depth 0pt width 0pt\vfill%
\special{postscriptfile #3}}}



\def\figHardAmps{
\begin{figure}[tbh]
 \epsfxsize=\hsize
  \epsfbox{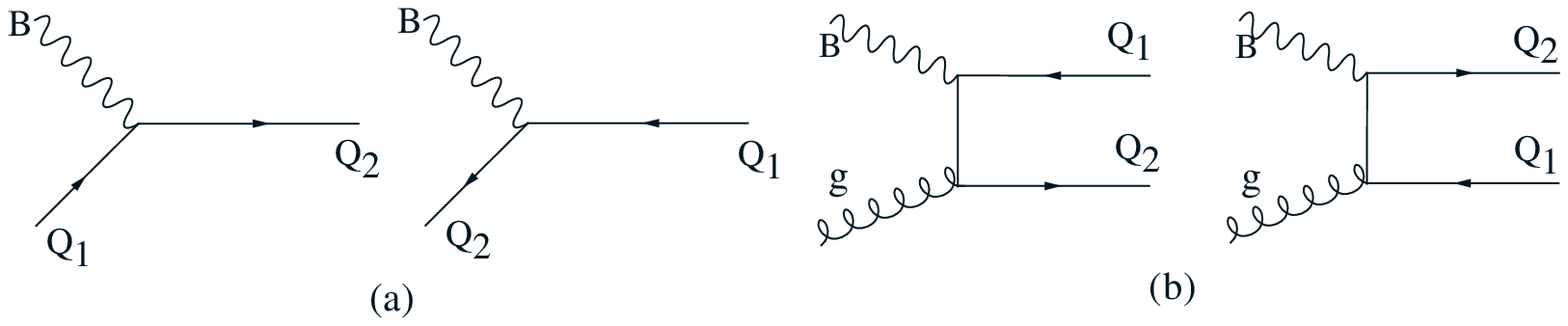}
      \caption{Amplitudes for heavy quark
      production: (a) order $\alpha _{s}^{0}$ quark-scattering; and (b) order $\alpha _{s}^{1}$
      gluon-fusion contributions.  At least one of the quarks, say $Q_{2}$, is ``heavy''
      and corresponds to the ``Q'' used in the text.  For flavor changing currents,
      the two quarks $Q_{1}$ and $Q_{2}$ are different. For neutral currents, they are the
      same.}
   \label{fig:HardAmps}
\end{figure}
}
\def\figxQplane{
\begin{figure}[tbh]
\epsfxsize=\hsize
  \epsfbox{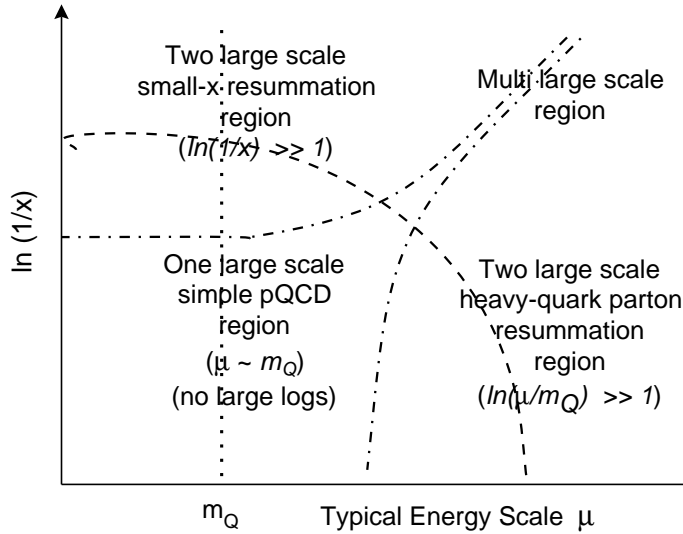}
 \caption{Regions of the $x-\mu _{{\it phy}}$ kinematic plane for a typical physical
   process involving a heavy quark with mass $m_{Q}$ and the natural QCD calculational
   schemes for each region.}
 \label{fig:xQplane}
\end{figure}
}
\def\figlhProc{
\begin{figure}[tbh]
  \epsfbox{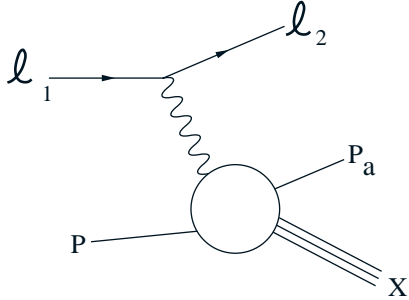}
 \caption{General lepton-hadron production amplitude for a heavy quark}
 \label{fig:lhProc}
\end{figure}
}
\def\figfact{
\begin{figure}[tbh]
 \epsfxsize=\hsize
  \epsfbox{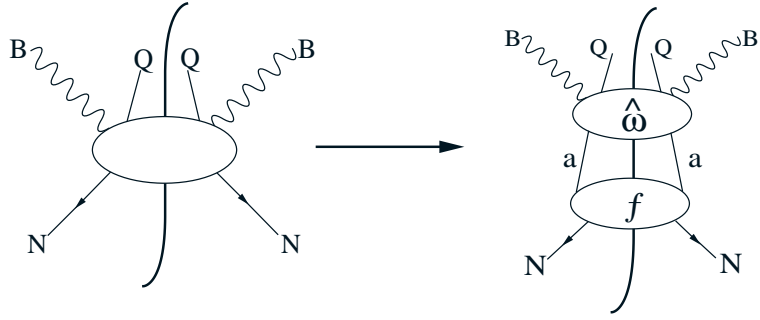}
 \caption{Graphical representation of the factorization formula,
  Eq.\ ({\protect \ref{HelGluFac}}).}
 \label{fig:fact}
\end{figure}
}
\def\figmasterEq{
\begin{figure}[tbh]
 \epsfxsize=\hsize
  \epsfbox{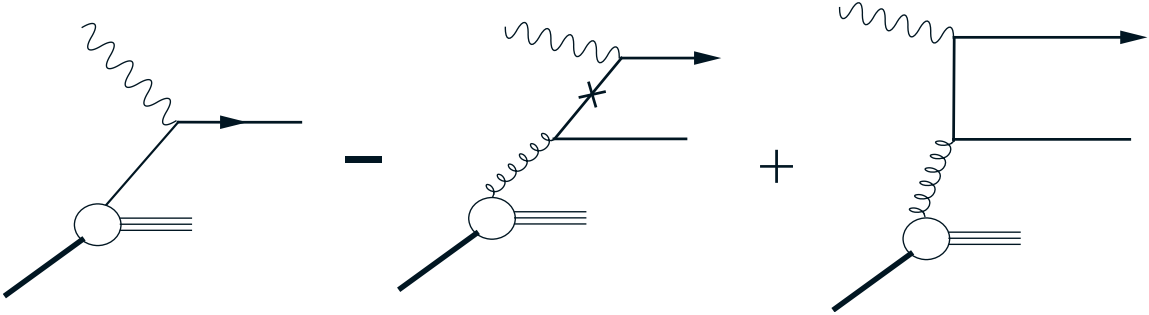}
 \caption{Graphical representation of the three terms which enter the
  {\em master equation}, Eq.\ ({\protect \ref{masterEq}}),
   for the physical structure functions: the subtraction term is placed in the middle to
  emphasize its similarity both to the quark-scattering (left) and to the gluon-fusion (right)
  contributions.
  The $\times $ on the internal quark line in the subtraction term indicates it is close to
  mass-shell and collinear to the gluon and the hadron momenta.}
 \label{fig:masterEq}
\end{figure}
}
\def\figBgSc{
\begin{figure}[tbh]
 \epsfxsize=\hsize
  \epsfbox{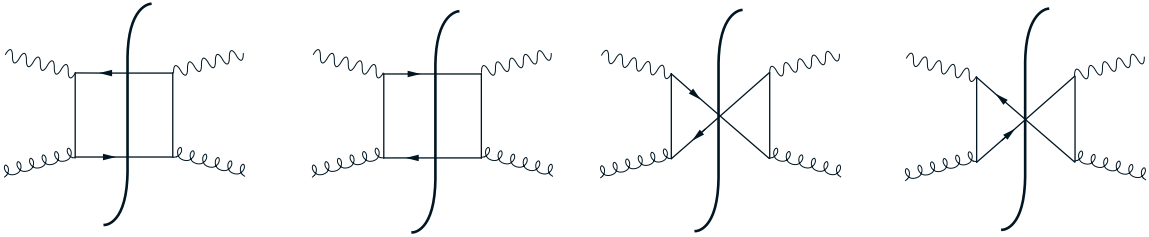}
 \caption{Cut diagrams for order $\alpha _{s}^{1}$ vector-boson gluon scattering.}
 \label{fig:BgSc}
\end{figure}
}
\def\figMuDep{
\begin{figure}[tbh]
\epsfxsize=\hsize
  \epsfbox{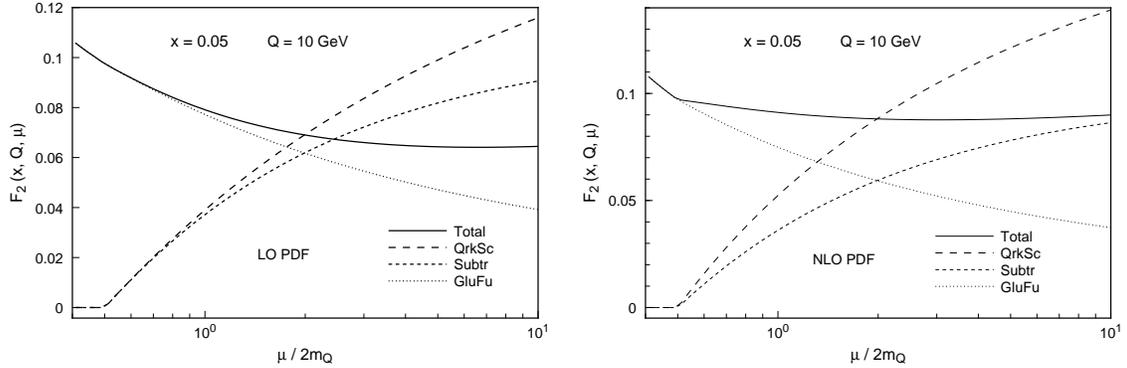}
 \caption{Scale dependence of the contributing terms to $F_{2}(x,Q)$.
  The factorization scale $\mu $ is shown in units of the physical scale
  $2 M_Q$.}
\label{fig:MuDep}
\end{figure}
}
\def\figLOqDep{
\begin{figure}[tbh]
\epsfxsize=\hsize
  \epsfbox{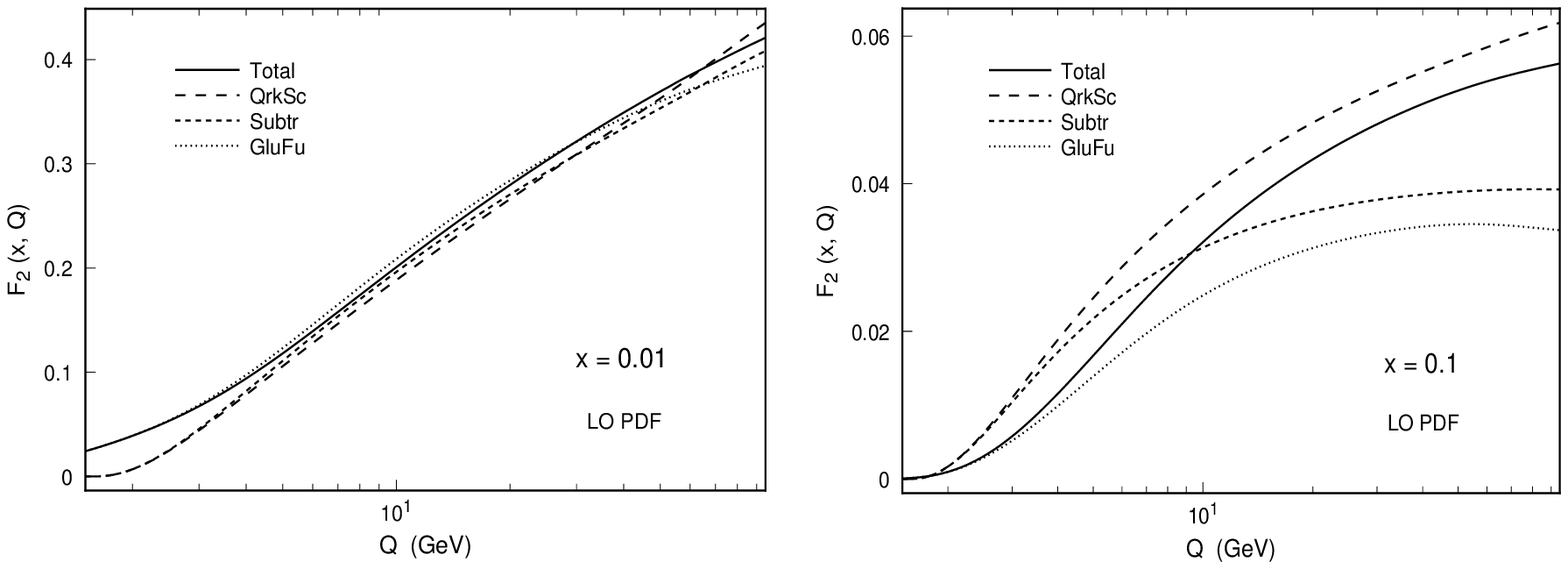}
 \caption{$Q$ dependence of $F_2(x,Q)$ at $x=0.01$ and $x=0.1$
 calculated using LO parton densities.
 }
\label{fig:LOqDep}
\end{figure}
}
\def\figNLOqDep{
\begin{figure}[tbh]
\epsfxsize=\hsize
  \epsfbox{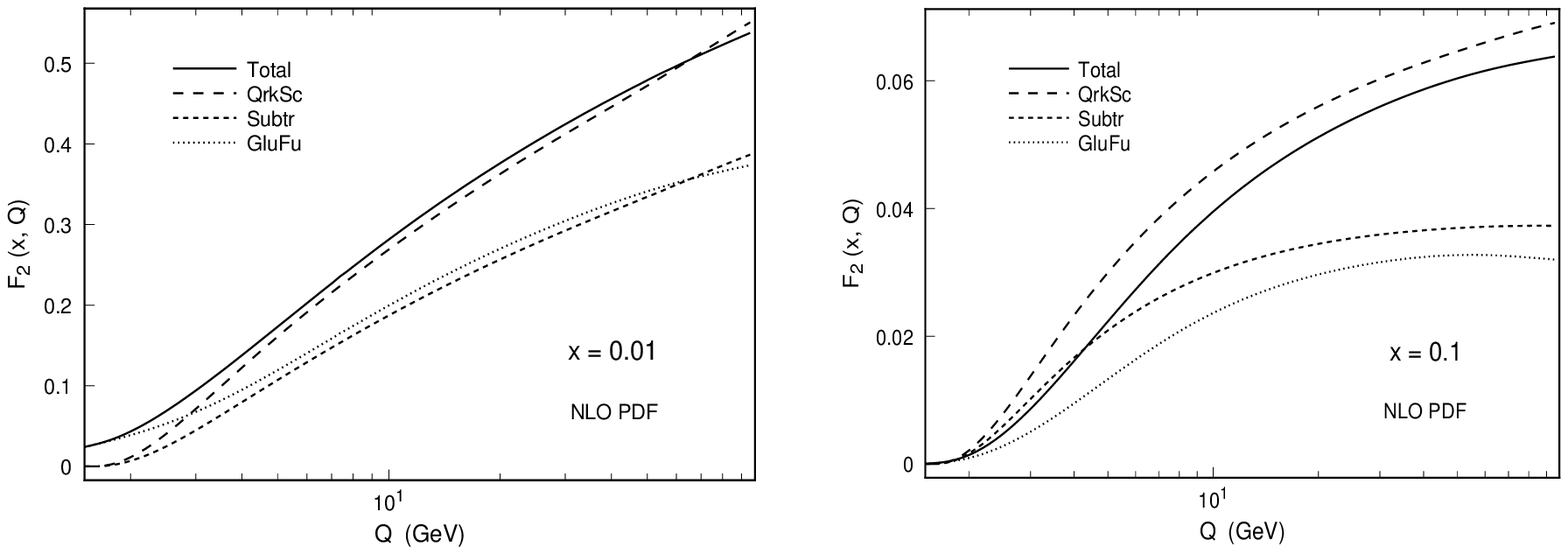}
 \caption{$Q$ dependence of $F_2(x,Q)$ at $x=0.01$ and $x=0.1$
 calculated using NLO parton densities.
 }
\label{fig:NLOqDep}
\end{figure}
}
\def\figxDep{
\begin{figure}[tbh]
 \epsfxsize=\hsize
  \epsfbox{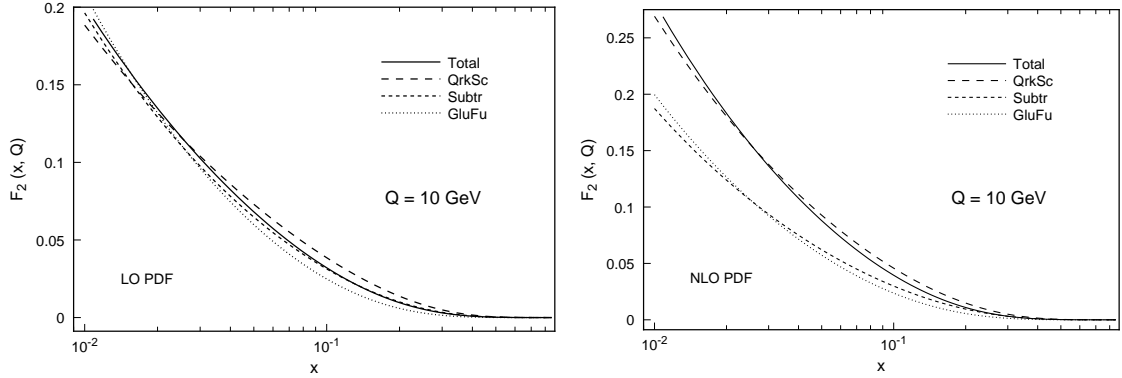}
 \caption{$x$ dependence of $F_2(x,Q)$ at $Q=10\;{\rm GeV}$
 calculated using LO and NLO parton densities.
 }
\label{fig:xDep}
\end{figure}
}
\def\figbProd{
\begin{figure}[tbh]
 \epsfxsize=\hsize
  \epsfbox{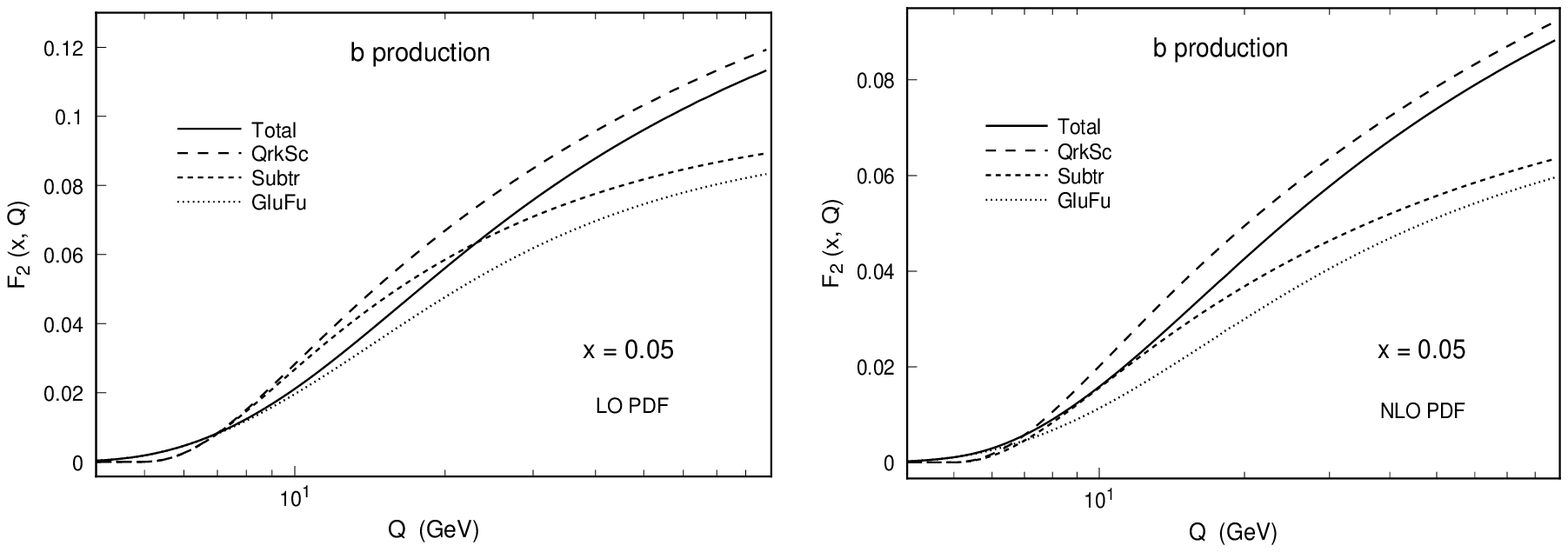}
 \caption{Structure function $F_{2}(x,Q)$ for b-production vs. Q at $x=0.01$.}
 \label{fig:bProd}
\end{figure}
}

\epsfverbosetrue

\null
\vfil

\begin{center}
\begin{tabular}{l}
October 1993
\end{tabular}
    \hfill
\begin{tabular}{l}
hep-ph/9312319 \\
SMU-HEP/93-17 \\
MSU-HEP 93/17 \\
PSU/TH/138  \\
\end{tabular}
\\[1cm]

{\LARGE Leptoproduction of Heavy Quarks II} \\
\vspace{1em}
{\Large -- A Unified QCD Formulation of Charged and Neutral
\\[0.1in]
Current Processes from Fixed-target to Collider Energies }
\\[.5in]

 {\large
   M. A. G. Aivazis,{${}^{a}$}
   John C. Collins,{${}^{b}$}
    Fredrick I. Olness,{${}^{a}$}\footnote{SSC Fellow}
   and Wu-Ki Tung${}^{c}$
  }
 \\[0.5in]
  {${}^{a}$}Southern Methodist University,
   Dallas, Texas  75275
 \\
  {${}^{b}$}Penn State University, University Park, PA 16802
 \\
  {${}^{c}$}Michigan State University
   East Lansing, Michigan  48824
\end{center}
\vfil

\begin{abstract}
A unified QCD formulation of leptoproduction of massive quarks in charged
current and neutral current processes is described. This involves adopting
consistent factorization and
renormalization schemes which encompass both
vector-boson-gluon-fusion (``flavor creation'') and
vector-boson-massive-quark-scattering (``flavor excitation'') production
mechanisms.  It provides a framework which is valid from the
threshold for producing the massive quark (where gluon-fusion is dominant)
to the very high energy regime when the typical energy scale $\mu $
is much larger than the quark mass $m_{Q}$ (where the quark-scattering should
be prevalent). This approach effectively resums all large logarithms of the
type $(\alfaS(\mu ) \log(\mu ^{2}/m_{Q}^{2}))^{n}$ which limit the validity of existing
fixed-order calculations to the region $\mu \sim O(m_{Q})$. We show that the
(massive) quark-scattering contribution (after subtraction of overlaps) is
important in most parts of the $(x, Q)$ plane except near the threshold
region. We demonstrate that the factorization scale dependence of the
structure functions calculated in this approach is substantially less than
those obtained in the fixed-order calculations, as one would expect from a
more consistent formulation.
\end{abstract}
 \vfil 


\newpage



\LaTeXparent{lhk2.tex}

\section{Introduction \label{sec:Intro}}

The production of heavy quarks in photo-, lepto-, and hadro-production
processes has become an increasingly important subject of study from both
the theoretical and experimental points of view. However, there are some
outstanding problems with existing perturbative QCD calculations of heavy
quark production: sizable (spurious) scale dependence of the predictions,
apparent disagreement with observed b-production cross-section at the
Tevatron, ... etc. See Ref.\cite{SmiTun93} for a recent review of the theory
and phenomenology of heavy quark production. As will be discussed later in
this paper, there is also an inconsistency in most theoretical calculations
of the cross sections: the schemes used in the next-to-leading order
calculations are not the same as those used in determining the parton
densities from global analyses, such as in
Refs.~\cite{MorTun,mrs92a,CTEQ1}. In
this paper we will spell out the details of a more complete and consistent
formulation of heavy quark production. For the sake of clarity, we shall
focus on the case of leptoproduction, although the same
principles apply to hadroproduction as well.

Perturbative QCD calculations rely on factorization
theorems$^{\cite{CSS86}}$:
Different factors involve different scales of virtuality, and a factor
that involves only physics on a scale $m$ can be effectively calculated in a
power series in $\alpha _{s}(m)$. The simplest factorizations, like the
operator product expansion, are for certain two-scale problems: One scale,
about the bound-state nature of hadrons, is of the order of $\Lambda $ or
the mass of a typical hadron; and the other, defining a scale of large
virtuality, can be $Q^{2}$ in deep-inelastic scattering or the transverse
momentum of a measured jet.\footnote{%
To avoid circumlocutions, we will often use the terminology of the operator
product expansion when discussing factorization theorems. In particular, we
will use the term ``Wilson coefficient'' to denote the short-distance
coefficient in the standard factorization theorem for deep inelastic
scattering etc.}

Processes involving heavy quarks are a good example of a multi-scale
process, for in practice we may then have to deal with at least four scales
(which we denote by $\mu _{i}$): $\Lambda $, $Q^{2}$ (as above),
and the masses of the charm and
bottom quarks, $m_{c}$ and $m_{b}$. When one uses conventional calculational
schemes designed for two-scale problems, the presence of more than one large
scale results in logarithms $\ln (\mu _{i}/\mu _{j})$ of the large ratios in the
higher order correction terms. These logarithms vitiate the very basis of
the original perturbative calculation, because of the large size of the yet
higher order terms beyond the order included. In view of the high energies
at present colliders, this problem can defeat the large effort put into
existing fixed order calculations.$^{\cite{SmiTun93}}$

The basic principles for treating this situation were constructed a long
time ago by Witten in his work$^{\cite{Witten}}$ on heavy quarks in the
operator product expansion (OPE). But his methods as they stand do not
provide a sufficient algorithm for calculating the processes in which we are
interested. An important practical concern is that the methods of
calculation should be applicable when some of the scales of interest are
comparable to each other as well as when they are very different. Simple
minded methods involving ``integrating out the heavy quarks'' are not
sufficient, but, as Collins, Wilczek and Zee$^{\cite{cwz}}$ pointed out a
long time ago, the problem can be conveniently considered as one of choice
of the subtraction scheme (for renormalization and factorization). We shall
refer to this work as CWZ in the following.

Consider the case of deep-inelastic scattering, with $Q$ denoting the
invariant mass of the exchanged boson, and with one heavy quark, of mass $m$%
. Three kinematic regions are of interest:

\begin{itemize}
\item  $m\gg Q$: The quark mass is larger than all other
scales in the problem, so
that the decoupling theorem$^{\cite{decoupl}}$ applies: All graphs involving
the heavy quark may be dropped, at the price only of a possible finite
renormalization of the parameters of the theory, notably $\alpha _{s}$.

\item  $m=O(Q)$: The heavy quark mass must be treated in the same way as $Q$
--- as a large parameter. Heavy quark lines appear in the Wilson coefficients
or in finite renormalizations of exactly the same kind as in the decoupling
theorem.

\item  $m\ll Q$: As far as the OPE is concerned, the heavy quark is to be
treated as light: its mass is to be neglected in Wilson coefficients, and
there are parton densities for the heavy quark. Since
the quark is heavy on an absolute scale,
{\it i.e.}, $m\gg \Lambda $, Witten's methods may be
used to calculate its density in terms of the densities of light partons.
\end{itemize}

The method we will describe will give a unified treatment that will cover
all ranges of mass. Furthermore, the CWZ method also allows us to treat the
case that there are several heavy quarks, whose masses may or may not be
strongly ordered.

\figHardAmps

One can see the issues in contrasting treatments of heavy quark production
in lepton-hadron scattering in existing literature.${}^{\cite{datareview}}$
For charged current interactions, such as charm production in neutrino
processes, most existing work focuses on the dominant underlying order-$%
\alpha _{s}^{0}$ parton process $W+s\to c$ ({\it cf.}\ Fig.~\ref{fig:HardAmps}%
a, the quark-scattering or ``flavor-excitation'' sub-process).$^{\cite {CCFR,FMMF,CDHSW,Gottschalk,bij,slowrescale,Brock}}$ In contrast, for
neutral current interactions, such as charm and bottom production in
electron- and muon-hadron scattering, practically all calculations begin
with the order-$\alpha _{s}^{1}$ parton process $\gamma /Z^{0}+g\to c+\bar c$
({\it cf}. Fig.~\ref{fig:HardAmps}b, the gluon-fusion or ``flavor-creation''
sub-process).$^{\cite{EMC,CHARM}}$ In both cases the initiating parton is a
light parton, and an appropriate scheme for computing radiative corrections
is what we will call a 3-flavor scheme, where the heavy quark only occurs in
the Wilson coefficients, and where there is no parton density for the charm
and bottom quarks.

But at very high energies (such as are now available at the HERA $e-p$
collider and beyond) charm and bottom quark masses can become small compared
to a typical energy scale: we can have $Q^{2}\gg m_{c}^{2},m_{b}^{2}$. It is then
natural to count these heavy quarks, especially the charm quark, as partons.$%
^{\cite{ali,schuler,reya,lampe}}$ We then effectively use a 4-flavor or
5-flavor scheme (as is done in commonly used parton distributions$^{\cite {MorTun,mrs92a,CTEQ1,tungowens}}$). In particular, the lowest order Wilson
coefficient for charm production in the neutral current process is not gluon
fusion, but flavor excitation $\gamma /Z^{0}+c\to c$. Of course, the gluon
density is rather large compared to the charm quark density, so that higher
order gluon fusion process can be numerically comparable or larger than the
flavor excitation process. (It is misleading to argue that an order $\alpha 
_{s}^{1}$ subprocess is smaller than an order $\alpha _{s}^{0}$ subprocess merely by
virtue of its being higher order unless the initiating partons are the
same.) Moreover, one must make the correct subtraction from the gluon fusion
process to avoid double counting.

Evidently, for this purpose, the notion of a quark with mass $m_{Q}$ being
``heavy'' or ``light'' must be taken as {\em relative} --- with respect to
the energy scale of the probe $\mu _{{\it phy}}$. The latter view forms the
very basis of the QCD parton model for the well-known light quarks $u,\ d$
and $s$ (which do have non-zero, albeit small, masses!). The parton approach
effectively resums large logarithms in fixed-order calculations arising from
initial state collinear singularities of the form $(\alpha _{s}\log \mu _{ {\it phy}}/m_{Q})^{n}$ to all orders in $\alpha _{s}$.

On the other hand, one also has an {\em absolute} notion for the term
``heavy quark'' ---that its mass is sufficiently large compared to $\Lambda $
so that $\alpha _{s}(m_{Q})$, the effective coupling at the heavy quark mass, is
in the perturbative region. This notion then refers to the charm, bottom and
top quarks as heavy, {\em regardless of the magnitude of the typical energy
scale ($\mu _{{\it phy}}$) of the problem}. This view is taken in all
next-to-leading order calculations on heavy quark production in the existing
literature.$^{\cite{SmiTun93,NDE,SmithEtAl}}$

As described earlier, the theoretical basis for a unified QCD treatment of
heavy quark in the {\em relative} sense, suitable for all energy scales,
already exists in the literature: it is based on the CWZ renormalization
scheme which naturally implements the intuitive energy-scale-dependent {\em %
light} and {\em heavy} quark concepts.$^{\cite{cwz,ColTun}}$ This scheme has
been applied with good success to the calculation of Higgs boson via heavy
quarks and gluons, effectively unifying the corresponding quark-scattering
and gluon-fusion subprocesses in one consistent scheme which is valid at all
energies.$^{\cite{OlnTun}}$ This approach clearly also provides a natural
framework for calculating the production of heavy quarks. It is particularly
simple to implement in lepto-production processes, as already pointed out in
a previous short communication.$^{\cite{AOT90}}$

\figxQplane

The present paper presents details of the method and the main physics
results. For definiteness, we shall refer to this approach as the {\em %
variable ({\it i.e.}, scale-dependent) flavor number scheme}, in contrast to
the {\em fixed flavor number scheme} used in conventional calculations of
heavy quark production.$^{\cite{NDE,SmithEtAl}}$ Since charged current and
neutral current processes are treated in one uniform framework, we shall use
the most general couplings for the vector gauge boson to the leptons and the
quarks; and we shall keep the most general mass configurations for the quark
lines in the hard cross-section calculation. The resulting complexity in
kinematics, in the application of the factorization theorem of QCD, and in
the calculation of hard matrix elements can be effectively handled using the
helicity formalism. This aspect of the problem is formulated and presented
in a separate paper,$^{\cite{AOT93a}}$ hereafter referred to as I. Sec.~\ref
{sec:OverView} provides an overview of the scale-dependent parton flavor
number scheme. Sec.~\ref{sec:1loopGlu} gives the detailed results on the
order $\alpha _{s}$ gluon-fusion amplitudes. Sec.~\ref{sec:Subtraction}
discusses the subtraction procedure needed to make quark-scattering and
gluon fusion mechanisms consistently co-exist. Sec.~\ref{sec:Results}
presents the main physics results to show the relative importance and
interplay of the various contributions, as well as the reduced
scale-dependence of the predictions. Finally, Sec.~\ref{sec:Discussions}
recapitulates the theoretical issues and points to potential
applications.

Since our approach effectively resums all large logarithms of the type $%
(\alpha _{s}\log \mu _{{\it phy}}/m_{Q})^{n}$ which occur in
fixed-parton-flavor-number calculations, it naturally extends the range of
validity of the latter beyond the region $\mu _{{\it phy}}\sim O(m_{Q})$. It
does not, however, deal with the class of logarithms of the type $(\alpha 
_{s}\log s/\mu _{{\it phy}}^{2})^{n}$ which is associated with the ``small-$%
x $'' problem (typically, $x=\mu _{{\it phy}}/\sqrt {s}$). The latter has
been the subject of several recent studies and it requires an entirely
different method of resummation --
the so-called $k_{t}$-factorization.$^{ \cite{LRSS89,CE91a,CHetc}}$
These two approaches are compatible and
complementary: they both extend the region of applicability of the
perturbative QCD calculations, but to different regions of phase
space.\footnote{A unified treatment is a topic for the future.}
This
is illustrated schematically in a map of the $x-\mu $ kinematic plane, Fig.~%
\ref{fig:xQplane}. Broadly speaking, our approach is needed when the typical
energy scale $\mu _{{\it phy}}$ becomes large (compared to $m_{Q}$) for
not-too-small $x;$ and the $k_{t}$-factorization method is necessary for
very small-$x$ and moderate $\mu _{{\it phy}}$. What values of $x$ must be
considered as ``small'', and $\mu _{{\it phy}}/m_{Q}$ as ``large'', to
require these improvements are open questions with no easy theoretical
answers in perturbative QCD. ({\it cf.}, the similar question: for what
value of $Q$ should Bjorken scaling set in?) However, these questions can be
investigated phenomenologically by comparing numerical results from the
different approaches in their regions of overlap. Existing numerical studies
of the small-$x$ resummation and conventional approaches suggest the latter
maybe valid down to $x\sim 10^{-4}$.$^{\cite{DESY90}}$ The results presented
later in this paper will shed some light on the comparison of
scale-dependent and fixed parton flavor number schemes.


\LaTeXparent{lhk2.tex}

\section{Overview of the Scheme and the Calculation\label{sec:OverView}}

We consider a general lepton-hadron scattering process:
\begin{equation}
\label{lhProc}\ell _{1}(\ell _{1})+N(P)\longrightarrow \ell _{2}(\ell 
_{2})+Q(p_{Q})+X(P_{X}),
\end{equation}
to lowest order in the electro-weak interactions, as depicted in Fig.~\ref
{fig:lhProc}. In the final-state, we have required that there be a heavy
quark $Q$ of momentum $p_{Q}^{\mu }$.\footnote{%
We mean here a heavy quark in the absolute sense: $m_{Q}\gg\Lambda $, so
that $\alfaS(m_{Q})$ is small enough to be in the perturbative regime. The
heavy quark will be detected by its hadronization products.} We label the
exchanged vector boson ($\gamma $, W, or Z) by $B$ and its momentum by $q$.

\figlhProc

\subsection{Hadron Structure Functions and Factorization\label
{sec:factorization}}

After the calculable leptonic part of the cross section has been factored
out, we work with the hadronic process induced by the virtual vector boson $%
B $:
\begin{equation}
\label{BhProc}B(q)+N(P)\longrightarrow Q(p_{Q})+X(P_{X}),
\end{equation}
and the cross-section is expressed in terms of the hadronic tensor
\begin{equation}
\label{Wtensor}W^{\mu \nu }=\frac {1}{4\pi }
\mathrel{\mathop{\overline{\sum }}\limits_{X(P_{X}),spin}}
\langle P|J^{\mu  }|P_{Q},P_{X}\rangle 
(2\pi )^{4}{\delta ^{(4)}\left( P+q-P_{Q}-P_{X}\right) }
\langle P_{X},P_{Q}|J^{\nu \dagger }|P\rangle .
\end{equation}
where $\overline{{\sum }}$ denotes a sum over all hadronic stateq
containing the final-state quark $Q$ of momentum $p_{Q}^{\mu }$.

\figfact

The factorization theorem asserts that the hadronic tensor
has the form$^{ \cite{CSS86}}$
\begin{equation}
\label{facThm}W_{BN}^{\mu \nu }(q,P,...) = f_{N}^{a} \otimes \widehat \omega 
_{Ba}^{\mu \nu } = \sum _{a}\int {\frac {d\xi }{\xi }}\ f_{N}^{a}(\xi ,\mu 
)\ \widehat \omega _{Ba}^{\mu \nu }(q,k_{a},...,\alfaS (\mu )),
\end{equation}
to the leading power of $q^{2}$. Here $f_{N}^{a}(\xi ,\mu )$ is the
distribution function of parton $a$ in the hadron $N$, and
$\widehat \omega _{Ba}^{\mu \nu }(q,k_{a},...,\alfaS (\mu ))$ is the Wilson coefficient.
That is, $\widehat \omega _{Ba}^{\mu \nu }$ is the same kind of object as the
hadronic tensor Eq.~(\ref{Wtensor}) except that it is evaluated
on a parton target,
and that the long-distance contributions are subtracted off.
The proof of
the factorization theoqrem is to show that these long-distance
pieces are all
correctly taken account of by the factor of the parton density in Eq.~(\ref
{facThm}).

The scale $\mu $ is the renormalization and factorization scale.\footnote{%
For simplicity, we do not distinguish between the factorization scale $\mu _{ {\it fac}}$ and the renormalization scale $\mu _{{\it ren}}$; we set them
equal to the same value $\mu $.} Roughly speaking, $\mu $ sets the
separation between the parts of the process that we attribute to long- and
short-distance phenomena. The predictive power of the factorization theorem
Eq.~(\ref{facThm}) arises when we set $\mu $ to a value of the order of a
large physical scale in the problem, say $\mu \,.b,q=\mu _{fac}=\mu _{ren})\approx \mu _{phy}\approx \sqrt {-q^{2}}$.
Then the hard scattering (or Wilson) coefficient $%
\widehat{\omega }$ may usefully be expanded in powers of the small coupling $%
\alfaS(\mu )$.  The $\mu $ dependence
of the parton densities $f_{N}^{a}$ is given by the Altarelli-Parisi equation,
whose kernel is also perturbatively calculable in powers of $\alfaS(\mu )$

As explained in I$^{\cite{AOT93a}}$, in the presence of non-zero masses, it
is the helicity amplitudes which provide the simplest connection between the
physical (scalar) structure functions $W_{BN}$ and the corresponding
parton-level quantities $\widehat{\omega }_{Ba}$. The factorization formula
then reads:
\begin{equation}
\label{HelGluFac}W_{BN}^{\lambda }(Q^{2},q\cdot p)=f_{N}^{a}\otimes \widehat{\omega }%
_{Ba}^{\lambda }=\sum _{a}\int {\frac {d\xi }{\xi }}\ f_{N}^{a}(\xi ,\mu )\ \widehat{%
\omega }_{Ba}^{\lambda }\left( \frac {x}{\xi },\frac {\hat s}{\mu },\frac {m_{Q}}{\mu },%
\alfaS (\mu )\right) ,
\end{equation}
where
$$
W^{\lambda }\ =\ \epsilon _{\mu }^{(\lambda )*}(q,p)\cdot W^{\mu \nu }\cdot 
\epsilon _{\nu }^{(\lambda )}(q,p)\quad \text{ ;}\quad \widehat{\omega }%
^{\lambda }\ =\ \epsilon _{\mu }^{(\lambda )*}(q,k)\cdot \widehat{\omega }^{\mu  \nu }\cdot \epsilon _{\nu }^{(\lambda )}(q,k),
$$
and $\epsilon _{\nu }^{(\lambda )}(q,r)$ is the polarization vector of the
vector boson with momentum $q$ and helicity $\lambda \ (=+,0,-)$ defined
with respect to the reference vector $r$. (Since $r$ is different in the
definition of $W^{\lambda }$ and $\widehat{\omega }^{\lambda }$, the simple
relation Eq.~(\ref{HelGluFac}) is not an obvious consequence of Eq.\ (\ref
{facThm}); it follows only because the two reference momenta $p$ and $k$ are
collinear, \cf\ I.) Fig.~\ref{fig:fact} depicts the factorization formula,
Eq.~(\ref{HelGluFac}), in a familiar form.

\subsection{Masses and factorization schemes}

The conventional method of calculation of the short-distance coefficients in
this and other QCD processes is to set to zero the masses of internal lines
and external partons in graphs for the partonic subprocesses. Then the
resulting infra-red poles (in dimensional regularization) are subtracted
according the \Msbar\ scheme. It can be shown that this implies that the
ultra-violet divergences in the definitions of the parton densities are
renormalized by the \Msbar\ scheme also.

Setting quark masses to zero gives the leading term in an expansion of the
short-distance coefficient in powers of $m_{Q}/\mu _{phy}$. This is obviously
inapplicable for a heavy quark if we want to treat the region where $\mu 
_{phy}$ is not much greater then $m_{Q}$. However, it is perfectly sensible to
leave the heavy quark mass in the calculation of $\widehat{\omega }$. We
will later show how this works in a calculation, and we will verify that the
$m_{Q}\to 0$ limit of the coefficient agrees with the standard zero mass
calculation. The parton densities, including the one for the heavy quark,
will continue to be defined by the \Msbar\ scheme. In the case that the
heavy quark is the charm quark, we will call this the ``4-flavor scheme''.%
\footnote{%
By keeping $m_{Q}$ non-zero in the Wilson coefficient, the ``theoretical
inconsistency'' described by Gl\"uck et al.$^{\cite{GRS}}$ does not
occur in our approach.
See further discussions in Sec.~\ref{sec:UEmassSubtr}}

On the other hand, when $\mu _{phy}\ll m_{Q}$, one finds that there are large
logarithms of the heavy quark mass in all perturbative calculations. That
is, the 4-flavor scheme does not manifestly exhibit decoupling of the heavy
quark. One obvious possibility is to use off-shell momentum-space
subtractions (which exhibit explicit decoupling) instead of \Msbar. But
this makes for much more complicated calculations, especially because of the
complicated off-shell structure of the renormalization counterterms for
gauge-invariant operators (such as are used to define the parton
densities).  The method of CWZ$^{\cite{cwz}}$ offers a natural and simple way:
switch to a ``3-flavor scheme'' in this region.

Technically, the CWZ scheme is a hybrid of \Msbar\ for the light partons and
zero-momentum subtraction for graphs with a heavy quark line. The scheme has
the following advantages: It satisfies manifest decoupling, and
preserves gauge invariance. The evolution
equations for the coupling and the parton densities are the same as for QCD
with 3 flavors of quark and pure \Msbar\ subtractions. Calculations are
quite simple compared with the off-shell scheme. Finally, the charm quark
density is zero to the leading power in $\Lambda /m_{c}$, so that the charm
quark mass only appears in Wilson coefficients.

The 3-flavor scheme is appropriate when $\mu _{phy}$ is comparable to or
less than about $m_{c}$. When other heavy quarks are present, one defines a
series of schemes: 3-flavor, 4-flavor, 5-flavor etc. The $N$-flavor scheme
is defined to treat the first $N$ flavors of quark as light, and the
remainder as heavy. It is appropriate when the physical scale of the
process, $\mu _{phy}$ is above the mass of quark $N$ and below that of quark
$N+1$. We will call $N$ the number of {\em active} quarks.

The relation between schemes with different numbers of active flavors is
just a case of a transformation between different renormalization and
factorization schemes; and the matching conditions between the schemes have
been calculated$^{\cite{ColTun,Qian,Rod}}$. At the one-loop level in the
\Msbar\ scheme, these are$^{\cite{ColTun}}$ just that the coupling and
parton densities are the same in the two schemes at $\mu =m_{Q}$. Thus a
convenient way of implementing them is to use 3-flavor evolution below $\mu 
=m_{c}$, to use 4-flavor evolution above that point, with continuity at the
break point. In this scheme, the use of $\mu =m_{Q}$ rather than, say, $\mu 
=2m_{Q}$, is a matter of explicit calculation using \Msbar\ subtraction, and
is not a matter of arbitrary choice. There are higher order corrections to
the matching conditions. Two-loop matching has been calculated$^{\cite{Rod}}$
for the coupling, but not yet for the parton densities.

It is worth noting that existing NLO calculations of heavy quark production$%
^{\cite{NDE,SmithEtAl}}$ essentially use the 3-flavor scheme as described
above -- for all energies, irrespective of the order of magnitude of $\mu 
_{phy}.$

\subsection{Contributing Partons and Parton Distributions\label{sec:pdfs}}

We use the term ``variable flavor number scheme'' to denote the scheme just
described. It is implemented$^{\cite{ColTun}}$ by using \Msbar\ evolution
with a number of active flavors that changes as one crosses the boundaries $%
\mu =m_{Q}$, where $m_{Q}$ is the mass of a heavy quark (charm, bottom, etc.).
The MT$^{\cite{MorTun}}$ and CTEQ$^{\cite{CTEQ1}}$ parton densities are
defined using this method. Thus for a given scale $\mu $ for the parton
densities, all quarks with mass less than $\mu $ are treated as partons (and
thus have associated QCD-evolved parton distributions). For a quark $Q$ with
non-zero mass $m_{Q}\ (\gg \Lambda _{QCD})$, $f_{N}^{Q}(\xi ,\mu )$ vanishes when $%
\mu \leq m_{Q}$ ({\it i.e.}, all the heavy quark dynamics in this region is in
the Wilson coefficients). But when $\mu >m_{Q}$, $f_{N}^{Q}(\xi ,\mu )$ satisfies
the usual \Msbar\ QCD evolution equation (with massless kernel functions)
above threshold. Thus, there is no fixed restriction on the sum over parton
flavor label $a$ in the basic factorization formula Eq.~(\ref{HelGluFac}):
depending on the value of the relevant $\mu $ ($\sim \mu _{{\it phy}}$) of
the physical process, the correct number of quark flavors appropriate for
that energy scale will contribute.

This conceptual and calculational simplicity has an associated price. In the
region just above the quark mass ($\mu \sim m_{Q}$), defining a parton
distribution function for $Q$ with massless evolution kernel appears to be
somewhat artificial. Indeed the use of $f_{N}^{Q}(\xi ,\mu )$ in a lowest order
parton model formula for a cross section tends to overestimate the cross
section, because the parton density does not contain the physical threshold
behavior. The errors are compensated when one brings in higher order terms
in the Wilson coefficient, as we will see. Although both schemes are equally
correct, in principal, it would seem better to use a fixed parton flavor
number around threshold, {\it e.g.}, 3 flavors for charm production. But as
one goes higher in scale, one is genuinely in the overlap region, where the
3-flavor and 4-flavor schemes are equally valid. Eventually, the 4-flavor
scheme becomes the one which describes the underlying physics more
accurately.

\subsection{Parton Structure Functions and Hard Scattering Mechanisms \label
{sec:HrdSc}}

In the variable flavor number scheme, for both charged current and neutral
current production of heavy quarks, initial state quark-partons contribute
through the vector-boson quark scattering (flavor excitation) subprocess,
Fig.~\ref{fig:HardAmps}a and its higher-order corrections; whereas the
gluon-parton contributes through the vector-boson gluon fusion (flavor
creation) subprocess, Fig.~\ref{fig:HardAmps}b, and its higher-order
corrections. The order $\alpha _{s}^{0}$ quark scattering hard amplitude $\omega 
_{BQ}^{\lambda (0)}$\thinspace is easy to calculate. Since it is obtained
from a simple tree diagram, we can identify it with the corresponding hard
amplitude $\widehat{\omega }_{BQ}^{\lambda (0)}$ which enters the
factorization formula, Eq.~(\ref{facThm}). In our framework, the explicit
results are given in I (Sec.~5 and Appendix C). The order $\alpha _{s}^{1}$
parton level gluon fusion amplitudes $\omega _{Bg}^{\lambda (1)}$ are also
relatively straightforward to evaluate since they are free from
singularities when all the quark masses are kept finite.\footnote{%
In the conventional calculation of photo- and lepto-production of heavy
quarks in the fixed parton flavor number scheme$^{\cite{NDE,SmithEtAl}}$
these are sometimes called Born terms because they represent the leading
order contribution to the flavor creation mechanisms. Since the flavor
excitation mechanism actually come in with one less power of $\alpha _{s}$,
we shall state the explicit powers of $\alpha _{s}$ to avoid confusion.}

Note the distinction between the notations $\widehat{\omega }$ and $\omega $
for the structure functions for the partonic subprocesses. {\em The
unadorned }$\omega ${\em \ will be a partonic structure
function\footnote{%
   We make a clear distinction between the concepts of `structure
   function' and `parton density', contrary to common usage.
}
without
subtractions, but with non-zero quark masses. }In the zero mass limit $%
\omega $ will be divergent.{\em \ {\em The hatted quantity }$\widehat{\omega 
}${\em \ will have subtractions to remove the infra-red dependence.} }It is
the subtracted partonic structure function {\em $\widehat{\omega }$ }that is
to be used in the factorization theorem Eq.~(\ref{facThm}).

We will present the detailed formulas for $\omega $ and $\widehat{\omega }$,
in the helicity formalism we use, later in Sec.~\ref{sec:1loopGlu}. Here we
focus on the relation between the unsubtracted $\omega $ and the subtracted $%
\widehat{\omega }$ at order $\alfaS$ in order to elucidate the underlying
principles. As is well known, this relationship is established by applying
the factorization formula at the parton amplitude level. This provides the
exact relation, to this order:
\begin{equation}
\label{facPar}
\begin{array}[t]{ll}
\omega _{Bg}^{\lambda (1)} & =\sum _{a}f_{g}^{a}\otimes
\widehat{\omega }_{Ba}^{\lambda }=\sum _{a}(f_{g}^{a(0)}\otimes \widehat{\omega }%
_{Ba}^{\lambda (1)}+f_{g}^{a(1)}\otimes \widehat{\omega }_{Ba}^{\lambda (0)})
\\  & =\widehat{\omega }_{Bg}^{\lambda (1)}+f_{g}^{Q(1)}\otimes \omega 
_{BQ}^{\lambda (0)},
\end{array}
\end{equation}
where we made use of the fact that $f_{b}^{a(0)}(\xi )=\delta _{b}^{a}\delta 
(1-\xi )$, $\widehat{\omega }_{Ba}^{\lambda (0)}=\omega _{Ba}^{\lambda (0)}$%
, and $Bg$ scattering begins at order 1. The order $\alpha _{s}$ quark
distribution inside an on-shell gluon, $f_{g}^{Q(1)}$, is given by
\begin{equation}
\label{QrkInG}f_{g}^{Q(1)}=\frac {\alpha _{s}(\mu )}{2\pi }\ln \frac {\mu ^{2}}{m_{Q}^{2} }P_{gQ},
\end{equation}
where we have used the \Msbar\ prescription to renormalize the ultra-violet
divergences in the quark density. Since we have kept a non-zero quark mass,
there is no infra-red divergence. In this formula, $P_{gQ}$ is the familiar $%
g\rightarrow Q\bar Q$ splitting function
$P_{gQ}(\xi )=\frac {1}{2}(1-2\xi +2\xi ^{2})$.
Eq.~(\ref{QrkInG}) follows from the Feynman rules for parton
densities, and it is quite accidental that there is no constant
term, but only the logarithm times the splitting function.
By inverting Eq.~(\ref{facPar}), we obtain the formula for the
requisite hard amplitude:
\begin{equation}
\label{nloHrd}\widehat{\omega }_{Bg}^{\lambda (1)}=\omega _{Bg}^{\lambda  (1)}\;-\;f_{g}^{Q(1)}\otimes \omega _{BQ}^{\lambda (0)}.
\end{equation}
The second term on the right-hand side (henceforth referred to as the {\em %
subtraction term}), represents that part of the gluon fusion term which at
large $\mu _{phy}^{2}$ has the internal quark line relatively close to the
mass-shell and almost collinear to the gluon.

\subsection{Complete NLO (order $\alpha _{s}$) Hadron Structure Functions%
\label{sec:masterEq}}

\figmasterEq

Combining these results, we obtain the formula for the physical helicity
structure functions for heavy quark production on a hadron target:
\begin{equation}
\label{masterEq}
W_{BN}^{\lambda }=f_{N}^{Q}\otimes \omega _{BQ}^{\lambda  (0)}\;
   -\;\sum _{i}f_{N}^{g}\otimes f_{g}^{Q_{i}(0)}\otimes \omega _{BQ_{i}}^{\lambda  (0)}\;
   +\;f_{N}^{g}\otimes \omega _{Bg}^{\lambda (1)}+O(\alfaS^{2}).
\end{equation}
Note that, in the case of a neutral current process,
a sum over the two heavy quarks (quark and antiquark)
in the final state is needed.
This is the basic QCD equation for lepto-production of heavy quark
production in our approach. The subtraction term is placed in the middle to
emphasize its similarity both to the quark-scattering (left) and to the
gluon-fusion (right) contributions. On one hand, this term overlaps with the
first (quark scattering) one due to the common factor $\omega _{BQ}^{\lambda  (0)}$ and the approximate equality
\begin{equation}
\label{approxFQ}f_{N}^{Q}(\xi ,\mu )\simeq f_{g}^{Q(1)}\otimes f_{N}^{g}\text{\ \ \
when }\mu >m_{Q}\text{ and }\alpha _{s}\,\log \frac {\mu }{m_{Q}}\sim O(1),
\end{equation}
where $f_{g}^{Q(1)}$ is given by Eq.~(\ref{QrkInG}).\footnote{%
It is straightforward to demonstrate that this expression satisfies the
leading order QCD evolution equation with the correct boundary condition
$f_{N}^{Q}(\xi ,\mu =m_{Q})=0$.} On the other hand, its close connection to
the last (gluon fusion) term originates from Eq.~(\ref{nloHrd}). Since it
represents the part of the gluon-fusion term which is already included in a
fully QCD-evolved quark-scattering term, a consistent formalism must lead to
its subtraction to avoid double counting, as naturally happens here. Fig.~%
\ref{fig:masterEq} illustrates the same point graphically. (For clarity, we
only show one-half of the cut-diagrams for this process, {\it
cf.}, Fig.~\ref{fig:fact}.)
The $\times $ on the internal quark line in the subtraction
term denotes the following operation:  In the hard scattering
part of the middle graph, that is, the upper part of the graph,
the incoming quark's momentum is replaced by an on-shell value
with zero transverse momentum.  This replacement gives a good
approximation when the quark is collinear to the gluon, and
results in a factor of the order $\alpha _{s}$ distribution of a quark
in a gluon, Eq.~(\ref{QrkInG}).

The physics behind this formula should be well-known to students of the
conventional QCD parton model for light quarks.\footnote{%
Although the nature of the subtraction term is usually not transparent to
most non-experts since, for zero-mass quarks, it is usually identified only
as the coefficient of a $1/\epsilon $ pole in the most commonly used \Msbar\
calculational scheme.} However, this formalism has not been invoked in
existing calculations of heavy quark production.$^{\cite {SmiTun93,NDE,SmithEtAl}}$ Rather, they typically use the scheme in which
the heavy quark only appears in the Wilson coefficients ({\it e.g.}, a
3-flavor scheme for charm production and a 4-flavor scheme for bottom
production). In that case, the quark-scattering (flavor-excitation)
contribution is excluded whenever $m_{Q}\neq 0$, no subtraction from the
gluon-fusion contribution is applied in those calculations. There are
calculations that allow the heavy quark to have a parton density, and thus
work well above the quark threshold. But these normally set the quark mass
to zero in the Wilson coefficients and thus are not good approximations when
the process is not sufficiently far above the quark threshold. ({\em Cf.}
Ref.\ \cite{GRS}.)

The variable parton-flavor-number scheme for calculating heavy quark
production thus represents a natural and correct
extension of the usual zero--quark-mass
QCD parton framework to the case of non-zero quark mass. (We avoid using the
word ``heavy'', at least temporarily, since at this point ``light'' and
``heavy'' are relative with respect to the typical energy scale in this
approach.) This scheme contains all the ingredients of a consistent QCD
theory of heavy quark production over a wide range of energy scales as
mentioned in Sec.~\ref{sec:Intro}. In particular, if the initial-state quark
(labeled by $Q$ in Eq.~(\ref{masterEq})) is massive and the typical energy
scale $\mu _{{\it phy}}$ is of the same order as $m_{Q}$, then Eq.~(\ref
{approxFQ}) implies an approximate cancellation of the first two terms on
the right-hand side of Eq.~(\ref{masterEq}) --- thus we arrive at $%
W_{BN}^{\lambda }\simeq \;f_{N}^{g}\otimes \omega _{Bg}^{\lambda (1)}$ --- {\it i.e.%
}, dominance of the gluon-fusion mechanism ---  which reproduces the usual
picture of heavy quark production in the fixed parton flavor number scheme.
(This is the region labeled ``one large scale $(m_{Q}\sim \mu _{phy})$'' in
Fig.~\ref{fig:xQplane}.)

On the other hand, if either $Q$ is a usual light quark ({\it i.e.}, $%
m_{Q}\approx 0$) or $Q$ is massive but $\mu _{{\it phy}}\gg m_{Q}$, Eq.~(\ref
{approxFQ}) does not hold; instead, the subtraction term becomes the
dominant piece of the gluon fusion contribution (because it has the large
logarithmic factor $\frac {\alpha _{s}(\mu )}{2\pi }\ln \frac {\mu ^{2}}{m_{Q}^{2}}$
embodying a ``collinear divergence''), hence the last two terms almost
cancel (leaving only a {\em correction term} of order $\frac {\alpha _{s}(\mu ) }{2\pi }$ with no large logarithm factor) and we obtain $W_{BN}^{\lambda }
=f_{N}^{Q}\otimes \omega _{BQ}^{\lambda (0)}+O(\alpha _{s})\;$ -- which reproduces
the leading order QCD parton model picture appropriate for energies much
higher than all masses. (This is the region labeled ``two large scales $%
(1\gg m_{Q}/\mu _{phy})$'' in Fig.~\ref{fig:xQplane}.)

Eq.~(\ref{masterEq}) provides a smooth interpolation between the two
kinematic regions described above, and contains both as special cases.
Because the subtraction term represents precisely the overlap of the other
two, a change in the factorization scale amounts explicitly to a reshuffling
between the three terms on the right hand side of Eq.~(\ref{masterEq}). The
differences arising from a change in the factorization scale are genuinely
of higher order in $\alpha _{s}$, hence are smaller by a factor $\alfaS$
than if one or more terms are left out.\footnote{%
   But remember that the gluon distribution is often much larger
   numerically than a quark distribution.
}
We will demonstrate this point in
detail in Sec.~\ref{sec:Results}. The differences can be made smaller by
using higher order terms in perturbation theory.

Strictly speaking, Eq.~(\ref{masterEq}) is incomplete: we should also add
order $\alpha _{s}$ quark-scattering contributions of the form $\;f_{N}^{Q}\otimes
\omega _{BQ}^{\lambda (1)}\;-\;f_{N}^{Q}\otimes f_{Q}^{Q^{\prime }(1)}\otimes
\omega _{BQ^{\prime }}^{\lambda (0)}$. The ideas are exactly the same as
discussed above for the order $\alpha _{s}$ gluon contributions; but these
terms are
numerically less important because we have many more gluons inside the
hadron then sea quarks. (In this sense, the order $\alpha _{s}$ $Bq$
scattering terms are {\em effectively one order higher} since, for sea
quarks, $f_{N}^{Q}\,\sim \;f_{N}^{g}\otimes f_{g}^{Q^{\prime }(1)}$ is of order $%
\alpha _{s}$ compared to the gluon distribution $f_{N}^{g}$.) We should also
mention that Eq.~(\ref{masterEq}) can be generalized to higher orders by the
systematic application of the above scheme. The order $\alpha _{s}^{2}$ hard
amplitudes will be given by formulas generalized from Eq.~(\ref{nloHrd}).
The differences between these and the corresponding ones already calculated
in the conventional fixed-number-of-flavor scheme are finite pieces
attributable to the change of renormalization scheme.


\LaTeXparent{lhk2.tex}

\section{One-loop Gluon-Initiated Parton Structure Functions\label
{sec:1loopGlu}}

The one-loop forward hard amplitudes for the $2\rightarrow 2$
vector-boson-gluon scattering process
\begin{equation}
\label{VecGluSc}B(q)+g(k)\rightarrow \overline{q}_{1}(p_{1})+q_{2}(p_{2})
\end{equation}
are given by the cut diagrams shown in Fig.~\ref{fig:BgSc}. As indicated, we
use $q$ to denote the momentum of the vector boson, $k$ the momentum of the
gluon in the initial state, and ($p_{1},$ $p_{2}$) the momenta of the quarks in
the final state.
For flavor changing charged current processes (vector boson $B=W^{\pm }$),
the subscript 1(2) will be associated with the light (heavy) quark; for
neutral current processes ($B=\gamma ,Z$), both are associated with the
heavy quark.

\figBgSc

Since we keep the masses of the quarks $(m_{1},m_{2})$ non-zero, these
diagrams do not contain any singularities: they are infra-red finite since
the gluon only appears in the initial state; and the internal
virtual quark lines
never go on mass-shell due to the finite masses. Thus it is safe to perform
the calculation in 4-dimensional space-time. The parton tensor
(cut) amplitudes
are represented by:
\begin{equation}
\label{TenGluAmp}\omega ^{\mu \nu }(q,k,p{_{1\,})\;=\;}\frac {1}{4\pi }\sum 
\nolimits_{i}\int d\Gamma _{2}\frac {N_{i}^{\mu \nu }}{P_{1}^{i}P_{2}^{i}},
\end{equation}
where $i$ labels the diagrams; $d\Gamma _{2}$ is the two-particle
final-state phase space differential; and $P_{1}^{i}P_{2}^{i}$ denotes the
two propagator factors appropriate for diagram $i$.

As mentioned earlier, we find it most convenient to work with helicity
structure functions. They are:
\begin{equation}
\label{HelGluAmp}
\begin{array}{rl}
\omega ^{\lambda }(Q{^{2},\hat s,m_{1}^{2},{m_{2}^{2}})\;=} & \dfrac{1}{4\pi } \sum \nolimits_{i}\int d\Gamma _{2}
\dfrac{N_{i}^{\lambda }}{P_{1}^{i}P_{2}^{i}} \\ N_{i}^{\lambda }\ = & \
\epsilon _{\mu }^{(\lambda )*}(q,k)\cdot N_{i}^{\mu \nu }\cdot \epsilon _{\nu }
^{(\lambda )}(q,k),
\end{array}
\end{equation}
where $\epsilon _{\nu }^{(\lambda )}(q,k)$ is the polarization vector of the
vector boson with helicity $\lambda \ (=+,0,-)$ defined with respect to $k.$
For the general case, the helicity amplitudes exhibit a natural symmetry if
we express the vector-boson-quark vertex in terms of the two chiral coupling
constants: $g_{R,L}$. The amplitudes $\omega ^{\lambda }$ (and $N_{i}^{\lambda }$)
are quadratic in the $g^{\prime }s$, and we can write
\begin{equation}
\label{ChiAmp}\omega ^{\lambda }=\frac {\alpha _{s}(\mu )}{\pi }\ \sum _{\kappa }\
C_{\kappa }\cdot \omega _{\kappa }^{\lambda }\quad \;\ (\kappa =s,\text{x},a)\,
\end{equation}
with the chiral coupling combinations (symmetric, anti-symmetric, and
crossed):%
$$
C_{s,a}=\ g_{R}^{2}\pm g_{L}^{2}\quad ;\quad C_{\text{x}}=2g_{R}g_{L}.
$$

It is convenient to express the results in terms of center-of-mass variables
of the parton process. We have
\begin{equation}
\label{CMmom}
\begin{array}{lcrccccr}
k^{\mu } & = & ( & \;\;k, & 0, & 0, & \;k & ) \\
q^{\mu } & = & ( & E_{q}, & 0, & 0, & -k & ) \\
p_{1}^{\mu } & = & ( & E_{1}, & \;p\sin \theta & 0, & \;p\cos \theta & ) \\
p_{2}^{\mu } & = & ( & E_{2}, & -p\sin \theta & 0, & -p\cos \theta & )
\end{array}
\end{equation}
where the initial state energy and momenta are%
$$
\begin{array}{lcl}
k & = & (\hat s+Q^{2})/(2 \sqrt {\hat s}) \\
E_{q} & = & (\hat s-Q^{2})/(2\sqrt {\hat s})
\end{array}
$$
and the final state variables are
\begin{equation}
\label{CMvar}
\begin{array}{lcr}
p & = & \Delta (\hat s,m_{2}^{2},m_{1}^{2})/(2 \sqrt {\hat s}) \\
E_{1} & = & (\hat s-m_{2}^{2}+m_{1}^{2})/(2 \sqrt {\hat s}) \\
E_{2} & = & (\hat s+m_{2}^{2}-m_{1}^{2})/(2\sqrt {\hat s})
\end{array}
\end{equation}
with $\Delta (a,b,c)\equiv \sqrt {a^{2}+b^{2}+c^{2}-2ab-2bc-2ca}$.

There are two types of logarithmic terms arising from the phase space
integration of the propagator factors in the $t$ and $u$ channels,
respectively:
\begin{equation}
\label{Glogs}
\begin{array}{rclcl}
L_{t} & = & \log \left( \dfrac{E_{1}+p}{E_{1}-p}\right) & \equiv & \log
\left[
\dfrac{(\hat s-m_{2}^{2}+m_{1}^{2}+\Delta (\hat s,m_{2}^{2},m_{1}^{2}))^{2}}{%
4m_{1}^{2}\hat s}\right], \\ [0.1in] L_{u} & = & \log \left( \dfrac{E_{2}+p}{%
E_{2}-p}\right) & \equiv & \log \left[ \dfrac{(\hat
s+m_{2}^{2}-m_{1}^{2}+\Delta (\hat s,m_{2}^{2},m_{1}^{2}))^{2}}{%
4m_{2}^{2}\hat s}\right].
\end{array}
\end{equation}

\subsection{General mass case (Flavor changing charge current interaction)
\label{sec:GenMas}}

For general masses $\{m_{1},m_{2}\}$, the independent right-handed helicity
structure functions are:
\begin{equation}
\label{Gright}
\begin{array}{lllll}
\omega _{s}^{+} & = & L_{t}\left[ \dfrac 12+\dfrac{E_{1}}k\left( \dfrac{E_{1}%
}k-1\right) \right] & - & \dfrac{2p}{\sqrt {\hat s}}\left( \dfrac{E_{q}}%
k\right) ^{2} \\  & + & L_{u}\left[ \dfrac 12+\dfrac{E_{2}}k\left( \dfrac{%
E_{2}}k-1\right) \right] &  &  \\
\omega _{x}^{+} & = & (L_{t}+L_{u})\dfrac{2m_{1}m_{2}}{(Q^{2}+\hat s)^{2}}%
(\hat s-(m_{2}^{2}+m_{1}^{2})) & - & \dfrac{8m_{1}m_{2}p\sqrt {\hat s}}{%
(Q^{2}+\hat s)^{2}} \\ \omega _{a}^{+} & = & L_{t}\left[ \dfrac 12+\dfrac{%
E_{1}}k\left( \dfrac{E_{1}}k-1\right) +\dfrac{m_{1}^{2}(m_{2}^{2}-m_{1}^{2})%
}{2\hat sk^{2}}\right] & + & \dfrac{p(m_{2}^{2}-m_{1}^{2})}{k^{2}\sqrt {\hat
s}} \\  & - & L_{u}\left[ \dfrac 12+\dfrac{E_{2}}k\left( \dfrac{E_{2}}%
k-1\right) +\dfrac{m_{2}^{2}(m_{1}^{2}-m_{2}^{2})}{2\hat sk^{2}}\right], &
&
\end{array}
\end{equation}
where the subscripts ($s,x,a$) refer to the chiral combinations of Eq.~(\ref
{ChiAmp}).

The longitudinal helicity ones are:
\begin{equation}
\label{Glong}
\begin{array}{lcl}
\omega _{s}^{0} & = & -(L_{t}+L_{u})\left[
\dfrac{(m_{2}^{2}+m_{1}^{2})(Q^{4}-(m_{2}^{2}-m_{1}^{2})^{2}-2\hat sk^{2})}{%
4\hat sQ^{2}k^{2}}\right. \\  &  & \left. \qquad \qquad -
\dfrac{E_{q}}{2k^{2}\sqrt {\hat s}}\left( m_{2}^{2}+m_{1}^{2}-\dfrac{%
(m_{2}^{2}-m_{1}^{2})^{2}}{Q^{2}}\right) +\dfrac{m_{2}^{2}m_{1}^{2}}{%
k^{2}\hat s}\right] \\  &  & -(L_{t}-L_{u})\,
\dfrac{E_{q}(m_{2}^{2}-m_{1}^{2})}{k^{2}\sqrt {\hat s}} \\  &  & +\dfrac
p{k^{2}Q^{2} \sqrt {\hat s}}\left[
(m_{2}^{2}-m_{1}^{2})^{2}+Q^{2}(2Q^{2}-(m_{2}^{2}+m_{1}^{2}))\right] \\
\omega _{x}^{0} & = & -(L_{t}+L_{u})\, m_{1}m_{2}\left[ \dfrac 1{Q^{2}}+
\dfrac{\hat s-(m_{2}^{2}+m_{1}^{2})}{2k^{2}\hat s}\right] +\dfrac{%
2m_{1}m_{2}p}{k^{2}\sqrt {\hat s}} \\ \omega _{a}^{0} & = & 0.
\end{array}
\end{equation}

And the left-handed helicity structure functions $\omega _{\kappa }^{-}$ are
related to the right-handed ones by the symmetry relations:
\begin{equation}
\label{Gleft}
\begin{array}{ccc}
\omega _{s}^{-}=\omega _{s}^{+},\qquad & \omega _{x}^{-}=\omega 
_{x}^{+},\qquad & \omega _{a}^{-}=-\omega _{a}^{+}.
\end{array}
\end{equation}

\subsection{Equal mass case (Flavor non-changing neutral current interaction)
}

When the masses are equal, $m_{1}=m_{2}=m$, we obtain, as a special case of the
above:%
$$
L_{t}=L_{u}=L=2\log \left[ \dfrac{\sqrt {\hat s}+\sqrt {\hat s-4m^{2}}}{2m}\right]
$$
$$
\omega _{a}^{0}=\omega _{a}^{-}=\omega _{a}^{+}=0
$$
and
\begin{equation}
\label{EqMas}
\begin{array}{l}
\omega _{s}^{+}=\omega _{s}^{-}=\ L\
\dfrac{(Q^{4}+\hat s^{2})}{(Q^{2}+\hat s)^{2}}\ -\ \dfrac{(\hat s-Q^{2})^{2}\Delta }{%
\hat s(Q^{2}+\hat s)^{2}} \\ \omega _{x}^{+}=\omega _{x}^{-}=\ L
\dfrac{4m^{2}(\hat s-2m^{2})}{(Q^{2}+\hat s)^{2}}\ -\ \dfrac{4m^{2}\Delta }{(Q^{2}+\hat
s)^{2}} \\ \omega _{s}^{0}=-L\
\dfrac{2m^{2}(4m^{2}Q^{2}+3Q^{4}-4Q^{2}\hat s-\hat s^{2})}{Q^{2}(Q^{2}+\hat s)^{2}}+\dfrac{%
4(Q^{2}-m^{2})\Delta }{(Q^{2}+\hat s)^{2}} \\ \omega _{x}^{0}=-L\ \dfrac{%
2m^{2}(-4m^{2}Q^{2}+Q^{4}+4Q^{2}\hat s+\hat s^{2})}{Q^{2}(Q^{2}+\hat s)^{2}}+\dfrac{%
4m^{2}\Delta }{(Q^{2}+\hat s)^{2}}
\end{array}
\end{equation}
where%
$$
\Delta \equiv \Delta (\hat s,m^{2},m^{2})=\sqrt {\hat s(\hat s-4m^{2})}.
$$


\LaTeXparent{lhk2.tex}

\section{Mass-singularities, Collinear Divergences, Subtractions and
Infrared-safe Amplitudes\label{sec:Subtraction}}

In the fixed parton-flavor-number calculational scheme, the results of the
last section represent the full answer to the vector-boson-gluon fusion
production of heavy quarks at the order $\alpha _{s}$ level. These results
contain terms which become large as one (or both) of the quark masses are
small compared to the characteristic energy scale. These {\em mass
singularity terms} are isolated by taking the $m_{1}$ ($m_{2}$) $\rightarrow \,0$
limits of Eq.~(\ref{Gright}-\ref{Gleft}). They arise from the configuration
in the phase space integration when an internal quark line becomes almost
on-shell\footnote{Relative to $Q^{2}$.}
and collinear to the initial state gluon. In our scheme with
scale-dependent parton-flavor-number, these terms are included in the quark
scattering contribution with properly evolved quark parton distributions. As
explained in Sec.~\ref{sec:masterEq} the QCD formalism provides a natural
procedure to subtract these terms from the gluon-fusion amplitudes to avoid
double counting. In this section, we identify the subtraction terms in some
detail.

\subsection{Unequal Mass (flavor-changing) Case ($m_{1}\rightarrow 0$) \label
{sec:UEmassSubtr}}

In the limit $m_{1}\rightarrow 0$, we drop all $m_{1}$ dependencies in Eq.~(\ref
{Gright}-\ref{Gleft}) except inside the logarithm where the mass singularity
resides. We also replace $m_{2}$ by $m_{Q}$ to emphasize that the remaining mass
is associated with a quark that is heavy in absolute terms ({\it i.e.},
compared to $\Lambda $). The arguments of the $t$- and $u$-channel
logarithmic factors become
\begin{equation}
\label{Log0m1}
\begin{array}{llcl}
L_{t}= & \log \dfrac{E_{1}+p}{E_{1}-p}=\log \dfrac{(E_{1}+p)^{2}}{m_{1}^{2}} & \rightarrow
& \log
\dfrac{4p^{2}}{m_{1}^{2}} \\ L_{u}: & \log \dfrac{E_{2}+p}{E_{2}-p}=\log \dfrac{(E_{2}+p)^{2}%
}{m_{2}^{2}} & \rightarrow & \log \dfrac{\hat s}{m_{Q}^{2}}
\end{array}
\end{equation}
where the magnitude of the quark 3-momenta is given by $p^{2}=(\hat
s-m_{Q}^{2})^{2}/(2\hat s)$. We note that $L_{t}$ is the only factor which contains
the {\em mass singularity} associated with $m_{1}\rightarrow 0$ . It arises
from the collinear integration region of the $t$-channel propagator factor
over the quark transverse momentum, Fig.~\ref{fig:BgSc}a,b,c. This is seen
as follows: in the limit under consideration, the singular factor in the
integrand of Eq.~(\ref{HelGluAmp}) is $1/(t-m_{1}^{2})\propto 1/(p_{t}^{2}+m_{1}^{2})$.
We can take the $m_{1}\rightarrow 0$ limit everywhere except in this factor
where it must be retained to cut off the collinear (i.e. $p_{t}=0$)
singularity. The leading behavior is obtained by keeping the constant term
of the Taylor expansion of the numerator function in $p_{t}^{2}$. We obtain
therefore,
\begin{equation}
\label{colinSin}\int _{0}^{p_{t\max }^{2}}\frac {dp_{t}^{2}}{p_{t}^{2}+m_{1}^{2}}=\log \dfrac{%
p_{t\max }^{2}}{m_{1}^{2}}=L_{t}+O(1)
\end{equation}
where we used $p=p_{t\max }$ for the quark lines, and explicitly displayed
the role of $m_{1}^{2}$ as the cutoff for the collinear singularity.

\vspace{2mm}\noindent{\bf Isolation of Infrared-sensitive terms:}

We can isolate the mass (collinear) singularity from the (process-dependent)
dynamics by writing
\begin{equation}
\label{massSin}L_{t}=\log \dfrac{\mu _{f}^{2}}{m_{1}^{2}}\,\;+\;\log \dfrac{%
4p_{t\max }^{2}}{\mu _{f}^{2}}
\end{equation}
where we have introduced an arbitrary scale parameter $\mu _{f}$ --- the
factorization scale --- which separates the low $p_{t}$ {\em collinear}
region from the true {\em hard scattering} region. For an appropriately
chosen $\mu _{f}$ ({\it e.g.}, some external physical scale {\em independent
of }$\hat s$), the first term on the right-hand side (the collinear term)
contains the mass singularity which is to be subtracted from the
gluon-fusion contribution (to make it infra-red safe) and resummed into the
QCD evolved quark distribution function in the quark-scattering contribution.

The infra-red sensitive terms can be collected by substituting Eq.~(\ref
{massSin}) in Eq.~(\ref{Gright}-\ref{Gleft}). We obtain the following
non-vanishing amplitudes:
\begin{equation}
\label{helIR}
\begin{array}{llccl}
\omega _{s}^{IR\;+} & = & 1 & \cdot & P_{g\to q}(\tilde x_{m})\log \left(
\dfrac{\mu _{f}^{2}}{m_{1}^{2}}\right) \\ \omega _{a}^{IR\;+} & = & 1 &
\cdot & P_{g\to q}(\tilde x_{m})\log \left(
\dfrac{\mu _{f}^{2}}{m_{1}^{2}}\right) \\ \omega _{s}^{IR\;0} & = & \dfrac{%
m_{Q}^{2}}{Q^{2}} & \cdot & P_{g\to q}(\tilde x_{m})\log \left( \dfrac{\mu 
_{f}^{2}}{m_{1}^{2}}\right)
\end{array}
\end{equation}
with $\tilde x_{m}$ denoting the scaling variable,
\begin{equation}
\label{xmscale}\tilde x_{m}=\tilde x\left( 1+\frac {m_{Q}^{2}}{Q^{2}}\right)
\;\;;\;\;\;\;\;\tilde x=\frac {Q^{2}}{2k\cdot q}=\frac {Q^{2}}{\hat s+Q^{2}}
\end{equation}
and $P_{g\to q}$ the usual gluon to quark splitting function,
\begin{equation}
\label{g2qSplit}P_{g\to q}(x)=\frac {1}{2}[(1-x)^{2}+x^{2}]
\end{equation}
They are exactly proportional to the leading order quark scattering
amplitudes in the same limit. Comparing Eq.~(\ref{helIR}) with the order $%
\alpha _{s}^{0}$ result from paper~I, we obtain:
\begin{equation}
\label{IR2Born}\omega _{Bg}^{IR\;\lambda }=\omega _{Bq}^{\lambda  (0)}(m_{1}=0)\cdot \frac {\alpha _{s}}{2\pi }P_{g\to q}(\tilde x_{m})\theta 
(\mu _{f}-m_{1})\log \dfrac{\mu _{f}^{2}}{m_{1}^{2}}\ \qquad \text{(for all }%
\lambda \text{)}
\end{equation}
This is anticipated in the discussion of the factorization theorem in Sec.~%
\ref{sec:OverView}, {\it cf.} the second term on the right-hand side of Eq.~(%
\ref{facPar}).

\vspace{2mm}\noindent{\bf Mass (Collinear) Subtraction and Infrared-safe
Amplitudes}:

As explained in Sec.~\ref{sec:OverView}, the collinear configuration from
which the above infrared sensitive amplitudes originate corresponds
physically to the overlapping region of the quark scattering and gluon
fusion production mechanisms. We need to subtract these amplitudes from $%
\omega _{Bg}^{\lambda }$ to avoid double counting and to obtain {\em infra-red
safe hard gluon scattering amplitudes} $\widehat{\omega }_{Bg}^{\lambda }$ (%
{\it cf.}\ Eq.~(\ref{nloHrd})). We have a certain freedom in choosing the
subtraction terms: as long as they contain the leading infrared sensitive
terms identified above, any specific choice made for these overlapping terms
effectively defines the factorization scheme inherent in the master
(factorization) equation, Eq.~(\ref{masterEq}). However, this freedom must
be constrained by two consistency requirements. On the one hand, the order $%
\alfaS^{0}$ quark amplitude $\omega _{Bq}^{\lambda (0)}$ should be identical
in its two occurrences in Eq.~(\ref{masterEq}). On the other hand, the
logarithmic term in Eq.~(\ref{IR2Born}) must agree with the scheme used to
define parton densities.

There is a certain amount of choice possible.

We consider the case of unequal masses, one large and one small:
$m_{2}\approx \mu _{f}=\mu _{phy}\gg m_{1}\gg \Lambda _{QCD}$.
Then the parton density that is needed for quark 1
is defined by \Msbar. Consider $\omega _{Bq_{1}}^{\lambda (0)}$, which is given by the
Born graph for scattering off the relatively light quark 1 to
make the heavy quark 2.  There are two
obvious choices for the value of $m_{1}$ in $\omega _{Bq}^{\lambda (0)}$: either
replace the mass $m_{1}$
of the lighter quark to zero, or leave it at the physical value. When $m_{1}$
is small compared with the physical scale for the hard
scattering, $\mu _{ {\it phy}}$, it certainly
does not matter what we do, since Wilson
coefficients have a finite zero-mass limit.
But our formalism also extends
to the region where $m_{1}$ is not negligible. Then the physics is correctly
given by the order $\alfaS$ graph for $Bg$ fusion, the third term $\omega 
_{Bg}^{\lambda (1)}$ in Eq.~(\ref{masterEq}) --- with the mass $m_{1}$ kept at
its physical value. (This is the region where the 3-flavor scheme applies
and it treats $m_{1}$ as ``heavy''.) For consistency, it is then advantageous
to keep $m_{1}$ also non-zero in the order $\alpha _{s}^{0}$ Wilson coefficient
$\omega _{Bq}^{\lambda (0)}$ even if this is not absolutely required by the
formalism --- the first two terms in the master equation are guaranteed to
cancel near threshold ({\it cf.}\ Sec.~\ref{sec:masterEq})
as long as the same
choice of $\omega _{Bq}^{\lambda (0)}$ is made in both terms.

Thus we shall define the subtraction term as
\begin{equation}
\label{SubtrGen}\omega _{Bg}^{Sub\;\lambda }=\omega _{Bq}^{\lambda  (0)}(Q^{2},m_{1},m_{Q})\cdot \frac {\alpha _{s}}{2\pi }P_{g\to q}(\tilde x_{m})\theta 
(\mu _{f}-m_{1})\log \dfrac{\mu _{f}^{2}}{m_{1}^{2}}\ \qquad \text{(for all }\lambda 
\text{)}.
\end{equation}
with the full mass dependencies in $\omega _{Bq}^{\lambda (0)}(\hat
s,m_{1},m_{Q})$, and with the understanding that the same choice of $\omega 
_{Bq}^{\lambda (0)}$ is to be made in the order $\alpha _{s}^{0}$
quark-scattering term in the master equation,
Eq.~(\ref{masterEq}). The choice of the \Msbar\ scheme for the
parton densities implies the precise formula given in
Eq.~(\ref{SubtrGen}), which has a logarithmic term, but no
constant term.  This follows from a calculation of the one-loop
density of a massive quark in an on-shell gluon.

The
properly subtracted, {\em infra-red safe hard gluon scattering amplitudes}
are then given by
\begin{equation}
\label{HrdAmpGen}\widehat{\omega }_{Bg}^{\lambda }(\hat s,Q^{2},m_{1},m_{Q},\mu 
_{f})=\omega _{Bg}^{\lambda }(\hat s,Q^{2},m_{1},m_{Q})-\omega _{Bg}^{Sub\;\lambda  }(Q^{2},m_{1},m_{Q},\mu _{f})
\end{equation}
with $\omega _{Bg}^{\lambda }(\hat s,Q^{2},m_{1},m_{Q})$ given by Eqs.~(\ref{Gright}%
)-(\ref{Gleft}) of Sec.~\ref{sec:GenMas} and $\omega _{Bg}^{Sub\;\lambda  }(Q^{2},m_{1},m_{Q},\mu _{f})$ by Eq.~(\ref{SubtrGen}) above respectively.

The explicit expressions for $\widehat \omega _{Bg}^{\lambda }$ in the
general case is not particularly illuminating. We give below the results in
the $m_{1}\rightarrow 0$ limit:
\begin{equation}
\label{Ghard1}
\begin{array}{lcl}
\widehat \omega _{s}^{+} & = & P_{g\to q}(\tilde x_{m})\log
\dfrac{4p_{t\max }^{2}}{\mu _{f}^{2}}-\ \dfrac{\tilde x(1-2\tilde x)^{2}}{%
(1-\tilde x)}\ \dfrac{(s-m_{Q}^{2})}{Q^{2}} \\  &  & +\;L_{u}\left[ P_{g\to  q}(\tilde x_{m})+2\tilde x(1-2\tilde x) \frac {m_{Q}^{2}}{Q^{2}}\right] \\
\widehat \omega _{a}^{+} & = & P_{g\to q}(\tilde x_{m})\log
\dfrac{4p_{t\max }^{2}}{\mu _{f}^{2}}+\ (2\tilde x^{2})\ \dfrac{%
m_{Q}^{2}(s-m_{Q}^{2})}{Q^{4}} \\  &  & -\;L_{u}\left[ P_{g\to q}(\tilde
x_{m})+2\tilde x(1-2\tilde x)
\dfrac{m_{Q}^{2}}{Q^{2}}-2\tilde x^{2}\left(
\dfrac{m_{Q}^{2}}{Q^{2}}
\right) ^{2}\right] \\
\widehat \omega _{\text{x}}^{+} & = & 0 \\
\widehat \omega _{s}^{0} & = & \frac {m_{Q}^{2}}{Q^{2}}P_{g\to q}(\tilde
x_{m})\log
\dfrac{4p_{t\max }^{2}}{\mu _{f}^{2}}+\ (2\tilde x^{2})\ \dfrac{%
(m_{Q}^{4}-m_{Q}^{2}Q^{2}+2Q^{4})(s-m_{Q}^{2})}{Q^{6}} \\  &  & +\;L_{u}
\frac {m_{Q}^{2}}{Q^{2}}\ \left[
\dfrac{1+6\tilde x-14\tilde x^{2}}2-\tilde x(1-2\tilde x)\frac { m_{Q}^{2}}{Q^{2}}+\tilde x^{2}\left( \frac {m_{Q}^{2}}{Q^{2}}\right)
^{2}\right] \\ \widehat \omega _{\text{x}}^{0} & = & \widehat \omega 
_{a}^{0}\ =\ 0
\end{array}
\end{equation}

Of course, for energy scales much larger then $m_{2}$ ($m_{Q}$), there is
also a mass singularity associated with the factor $\log (m_{2}/\mu _{f})$
which resides in $L_{u}$. In our approach, an infra-red sensitive term
analogous to Eq.~(\ref{IR2Born}) with $m_{1}\leftrightarrow m_{2}$ should
then be subtracted from the order $\alpha _{s}$ gluon-fusion amplitudes. The
subtracted part, again, is included in the corresponding order $\alpha 
_{s}^{0}$ quark scattering amplitude --- with incoming quark ``2'' --- which
represents the resummed result of all such terms to arbitrary orders.
The resummation is performed by the Altarelli-Parisi evolution of
the parton densities.

\subsection{Equal Mass Case ($m_{1}=m_{2}=m\rightarrow 0$) and Comparison
with $\overline{MS}$\ scheme results}

Corresponding results for the case of flavor non-changing neutral current
interactions can be obtained from the above general results by setting $%
m_{1}=m_{2}=m$ and choosing the appropriate couplings. We give a few
explicit formulas for illustrative purpose and for establishing the relation
of our subtraction scheme to the \Msbar\ scheme. For the equal mass case,
the infra-red sensitive logarithm factor is
$$
L=\log \frac {4p_{t\max }^{2}}{m^{2}}=\log \frac {\hat s}{m^{2}}=\log \frac { \mu _{f}^{2} }{m^{2}}+\log \frac {\hat s}{\mu _{f}^{2}}
$$
In the $m\rightarrow 0$ limit, the non-vanishing helicity amplitudes are
(keeping $m$ only in the other-wise divergent logarithm)
\begin{equation}
\label{0mass1}
\begin{array}{rcl}
\omega _{s}^{+}= & \omega _{s}^{-}= & 2P_{g\to q}(\tilde x)\log
\dfrac{\hat s}{m^{2}}-(1-2\tilde x)^{2} \\  & \omega _{s}^{0}= & 4(1-\tilde
x)\tilde x
\end{array}
\end{equation}
After subtracting the mass-singularity, we obtain the infra-red safe hard
amplitudes for zero mass quarks
\begin{equation}
\label{0mass2}
\begin{array}{rrl}
\widehat \omega _{s}^{+}= & \widehat \omega _{s}^{-}= & 2P_{g\to q}(\tilde
x)\ \log
\dfrac{\hat s}{\mu _{f}^{2}}-(1-2\tilde x)^{2} \\  & = & 2P_{g\to q}(\tilde
x)\ (\log
\dfrac{Q^{2}}{\mu _{f}^{2}}+\log \dfrac{1-\tilde x}{\tilde x})-(1-2\tilde
x)^{2} \\  & \widehat \omega _{s}^{0}= & 4(1-\tilde x)\tilde x
\end{array}
\end{equation}
These correspond to the ``Wilson coefficients'' for deep
inelastic scattering,
as usually calculated in the \Msbar\ scheme. It is
straightforward to verify
that they, indeed, are identical to the \Msbar\ Wilson
coefficients.${}^{ \cite{FurPet}}$
Hence, {\em our subtraction prescription (applicable to the
general mass case) reduces to the \Msbar-scheme of subtraction of collinear
singularities in the zero-mass limit}. This is simply a consequence of our
choice of the \Msbar\ scheme to define the parton densities, which then
resulted in Eq.~(\ref{QrkInG}) for the one-loop value of the quark density
in an on-shell gluon. A change of definition of the parton densities would
have added an infra-red safe term to Eq.~(\ref{QrkInG}), and there would be
corresponding terms to be added to the other terms in the
formulae we have written.


\LaTeXparent{lhk2.tex}

\section{Results on Structure Functions\label{sec:Results}}

We shall now study the numerical significance of the quark scattering (QS)
mechanism compared to gluon fusion (GF) in this unified framework. For the
case of charged current production of a heavy quark from a light quark, we
have demonstrated in a previous publication that the two basic processes are
of the same size numerically; hence a quantitative QCD analysis must
incorporate both in a consistent manner such as formulated above.$^{\cite {AOT90}}$ In the following, we shall concentrate on heavy quark production
by neutral current interaction which is of great interest both at fixed
target and ep collider energies. For the purpose of this paper, we shall
present results on the heavy quark production structure functions $F_{2}(x,Q)$
(which can be applied at all relevant energies). Phenomenological results on
cross-sections for various interesting processes at specific energies of
fixed target and HERA experiments will be pursued in a subsequent study. For
practical reason, we show mostly results on charm production. At the end, we
also show some corresponding results on b-production.

Two sets of parton distributions are used in the following study: the
``next-to-leading order (NLO) set'' consists of the
CTEQ2M distributions,$^{ \cite{CTEQ2}}$ and
the ``leading order (LO) set'' which is generated from
initial distributions at $Q_{0}=1.6\,$GeV taken from CTEQ2M but evolved using
LO evolution kernel only.
The differences between using LO and NLO are quite substantial,
as regards the charm-quark distribution, and this is a symptom
that higher order terms are important.

\subsection{Scale-dependence of the Structure Function\label{sec:ScDep}}

As with all applications of the perturbative QCD parton formalism, a
theoretical uncertainty about these calculations concerns the choice of
factorization scale (which we identify with the renormalization scale, {\it %
cf.}, Sec.~\ref{sec:Intro}). Whereas some scale dependence of the
theoretical prediction is unavoidable to any given order in $\alpha _{s},$ an
excessive sensitivity to the scale parameter usually signals a large
theoretical uncertainty. This has been a worry for existing NLO calculations
of charm and bottom production, especially for hadronic scattering.$^{\cite {NDE},\cite{ADMN88}}$ In order to make clear how the physical results
presented later might depend on the (implicit) choice of scale, we first
investigate the scale dependence of the various contributions which enter
our calculations. In particular, we demonstrate that whereas both the
quark-scattering (QS) and the gluon-fusion (GF) terms show substantial scale
dependence, these dependences are opposite in direction and they compensate
each other when combined according to the variable flavor number scheme,
{\em cf.} Eq.~(\ref{masterEq}).

\figMuDep

Fig.~\ref{fig:MuDep}a shows $F_{2}(x,Q;\mu )$ as a function of $\mu $ for $x=0.05$
and $Q=10$~GeV, calculated
using the LO parton distribution set. We display the $\mu $%
-dependence of the QS (long-dashed line), the GF (dotted line), and the
Subtraction (short-dashed line) terms individually along with the combined
(solid line) Total result. We see that the rapid rise of the QS and the
Subtraction terms together with the somewhat gentle fall of the GF
contribution combine to make the Total result substantially more stable then
either of the two individual production mechanisms. We also note the
following important features of Fig.~\ref{fig:MuDep}a: for $\mu $ below the
heavy quark mass, the QS and subtraction terms vanish by definition
({\it cf.}\ Sec.~\ref{sec:pdfs}
and Eq.~(\ref{approxFQ})), we have $F_{2}^{Tot}=F_{2}^{GF}$;
for $\mu $ just above the mass threshold, the QS and Subtraction
contributions nearly cancel according to Eq.~(\ref{approxFQ}),
we have $F_{2}^{Tot}\simeq F_{2}^{GF}$ (and the difference has reduced $\mu $
dependence); but for $\mu $ much larger than the mass threshold, QS (after
subtraction) makes a substantial contribution to $F_{2}^{Tot},$ and $F_{2}^{GF}$
ceases to be a good approximation$.\ $These features provide supporting
evidence to the theoretical discussion of Sec.~\ref{sec:factorization}-\ref
{sec:masterEq}, as will be reinforced by results to follow.

Fig.~\ref{fig:MuDep}b shows the
corresponding results obtained with the NLO parton
distribution set. All the
qualitative features are the same as above. However,
the cancellation between the QS and Subtraction terms above the heavy quark
threshold is not as complete as in Fig.~\ref{fig:MuDep}a.  (The
reason is that
Eq.~(\ref{approxFQ}) is not as good an
approximation as in the previous case, because the parton
densities are evolved at the NLO, but Eq.~(\ref{approxFQ}) is
only used at order $\alpha _{s}^{1}$ in our calculation.)
Hence
the contribution of the QS production mechanism (after subtraction) to the
structure function is even more significant for all values of $\mu $ above
the threshold. The NLO parton distribution functions contain resumed
sub-leading logarithms of $\mu /m_{Q}$, thus the difference between the QS and
Subtraction terms contains additional pieces of higher order terms not
present in the order $\alpha _{s}^{1}$ GF calculation included in this study.
Note the complete stability of the Total curve against the choice of $\mu $
in this case.

These results imply that, in subsequent discussions of the $x$- and $Q$%
-dependencies of $F_{2}(x,Q)$, the choice of the scale $\mu $ can shift the
QS, GF and Subtraction terms individually by considerable amount, but it
will not affect the Total answer by nearly as much. This fact underscores
the intrinsic inter-dependency of the two heavy quark production mechanisms
(as a basic quantum mechanical mixing effect). It also ensures that we
actually have a fair range of freedom of choice of the scale --- which we can
take advantage of whenever there is good physics reason to do so.

\subsection{Choice of Scale}

When $Q\gg m_{Q}$, the natural hard scale of the production process is of the
order $Q$. For $Q\sim O(m_{Q})$, $\mu $ can in principle be any combination of
$Q$ and $m_{Q}$ which is of the same order of magnitude$.$ To make an
intelligent choice however, it is important to be guided by relevant
physical considerations. Since we know that gluon fusion represents the
correct physics near the threshold for heavy quark production, it is
appropriate to choose a $\mu $ such that the quark scattering contribution
(along with the subtraction term) becomes small in this region. It also makes
sense to let the latter vanish when $Q<m_{Q}$. As a concrete example, the
following ansatz for the scale $\mu $ satisfies all these requirements.
\begin{equation}
\label{Qscale}
\begin{array}[t]{cll}
\mu ^{2} & =m_{Q}^{2}+c\,Q^{2}\,(1-\frac {m_{Q}^{2}}{Q^{2}})^{n} & \qquad
\text{for\quad }Q>m_{Q} \\  & =m_{Q}^{2} & \qquad \text{for\quad }Q\leq m_{Q}
\end{array}
\end{equation}
The results presented below are obtained using this ansatz with $c=0.5$ and $%
n=2$. Of course, as is the case with all pQCD calculations, an infinite
number of other choices are also acceptable. However, the results of the
last subsection ensure that, in our formalism, qualitative features of the
answer will be common to most reasonable choices --- as we have verified by
actual calculation with a variety of prescriptions for $\mu $. (This is
obviously not the case if the QS or GF production mechanisms are taken
individually, particularly if an extended range of $Q$ is involved.)

\subsection{Behavior of Structure Functions and the Interplay between the QS
and GF Production Mechanisms\label{sec:QSvsGF}}

Fig.~\ref{fig:LOqDep}a shows $F_{2}(x,Q)$ for charm production as a function of $Q$ for
fixed $x=0.01$ using LO parton distributions. The lines are labeled the
same way as in Fig.~\ref{fig:MuDep}.
For our particular choice of scale, the QS
contribution (long-dashed line) emerges from threshold and becomes
comparable in size to the GF contribution (dotted line) beyond around 5 GeV.
Because the subtraction term tracks the QS term rather closely throughout
the kinematic range except at the very large $Q$ end, the net contribution
of these two is quite negligible, hence the Total curve (solid line) stays
very close to the GF one except for very large $Q$.

\figLOqDep

Fig.~\ref{fig:LOqDep}b is analogous to Fig.~\ref{fig:LOqDep}a
except that we now look at the
behavior of the various contributions to $F_{2}(x,Q)$ at a larger $x$ value, $%
x=0.1$. The picture is somewhat different and, for the purpose of
illustrating the physics underlying our approach, more illuminating. For the
same choice of scale as above, the QS contribution rises rather steeply, and
overtakes the GF contribution almost immediately above threshold. Of special
interest is the behavior of the subtraction term (short-dashed line): it
tracks the QS contribution above threshold, as noted before, then turns to
follow the GF curve at large $Q$. The latter behavior follows from the
definition of the subtraction as the leading collinear log term of the
(order $\alpha _{s})$ GF contribution, {\it cf.} Sec.~\ref{sec:masterEq} and
Sec.~\ref{sec:UEmassSubtr}. As a result, {\em the complete QCD result (the
``Total'' curve) follows the GF term at low $Q$, but it approaches the QS
contribution at high $Q.$}
We note, this same behavior is also present in
Fig.~\ref{fig:LOqDep}a, for $x=0.01$.
However, there it is obscured by the closeness of the QS and GF terms due to
our particular choice of scale and parton distributions. 

\figNLOqDep

In Figs.~\ref{fig:NLOqDep}a
and \ref{fig:NLOqDep}b,
we show the same curves as in Figs.~\ref{fig:LOqDep}a and
\ref{fig:LOqDep}b
respectively, now calculated using NLO parton distributions. The same
physics effects are clearly displayed in both cases, only in more dramatic
proportions.

The general feature of {\em close interplay between QS and GF production
mechanisms} follows directly from our basic premises discussed in the
introductory sections, Sec.~\ref{sec:Intro} and Sec.~\ref{sec:OverView}:
when the relevant physical energy scale ($Q$) is comparable to the mass of
the heavy quark ($m_{Q}$), this quark behaves more like a heavy particle
rather than a parton, hence GF is the dominant production mechanism;%
\footnote{%
This is the basic tenet of the analysis of Collins, Soper and Sterman, Ref.~%
\cite{CSS86}. The underlying physics was quantitatively demonstrated
previously in a different process by Olness and Tung in Ref.~\cite{OlnTun}.}
but when $Q\gg m_{Q}$, it behaves characteristically like a light quark parton
almost by definition, and QS becomes the dominant process. This intuitively
reasonable behavior naturally emerges from the {\em variable
(scale-dependent) flavor-number scheme} of calculating massive quark
production, Eq.~(\ref{masterEq}).

\figxDep

We now look at $F_{2}(x,Q)$ as a function of $x$ at fixed $Q.$
Figs.~\ref{fig:xDep}a and b
show the various terms for $Q=10\;{\rm GeV}$ calculated
using LO and NLO parton distributions respectively. We see the
characteristic rise of the structure function toward small $x$. For fixed
$Q$, the large $x$ limit corresponds to the total final-state energy $%
W\rightarrow W_{{\it threshold}}$ (for heavy quark production); all
contributions become small. As $x\rightarrow 0$, one moves away from the
threshold region, the relative size of the various terms are sensitive to
the choice of scale and the choice of parton distribution functions.

These results clearly illustrate the importance of the QS (``flavor
excitation'' in old literature) mechanism for heavy flavor production. The
GF (``flavor creation'') mechanism provides a natural explanation of the
production of the heavy quarks not far above the threshold; but it is not
adequate to account for this process when the energy scale becomes large.
Part of the QS contribution (with proper subtraction) will be included when
the next order ({\it i.e.}, $\alpha _{s}^{2}$) GF hard-scattering terms are
included.\cite{SmithEtAl} However, the latter only contain terms to order $%
\alpha _{s}^{2}\ln {}^{2}(\mu /M_{Q})$, whereas the QS contribution represents the
resummed results of all such terms to arbitrary orders.

\subsection{B-quark Production}

Results on b-quark production are similar to those shown for charm. We show
only one plot, Fig.~\ref{fig:bProd},
which shows $F_{2}(x,Q)$ as a function of $Q$ for fixed $x=0.01
$. We see that the features are entirely similar to those seen in
Figs.~\ref{fig:LOqDep} and \ref{fig:NLOqDep} for charm production.

\figbProd


\LaTeXparent{lhk2.tex}

\section{Discussions \label{sec:Discussions}}

We have shown in this paper that the currently available fixed-order QCD
calculations of heavy quark production using the flavor creation mechanism
alone have to be generalized to include flavor excitation with
scale-dependent number of quark-flavors in order to account for the
appropriate underlying physics at all energies. A consistent scheme to
implement this generalization is formulated in detail to order $\alpha _{s}$
for leptoproduction here. The method can be extended to higher orders: in
addition to the $\alpha _{s}^{2}$ flavor-creation diagrams which are already
calculated in the literature, one needs to add order $\alpha _{s}$
flavor-excitation (vector-boson scattering off ``heavy quark'' partons)
contributions and perform the appropriate subtraction. As remarked earlier,
these two contributions are numerically comparable in spite of the formal
difference in the power of $\alpha _{s}$ by one. Exactly the same principles
apply to hadroproduction. As the results of the previous section show, the
inclusion of the right physics in these calculations can be expected to
improve the theoretical accuracy of QCD predictions --- as illustrated by
the reduced dependence on the choice of (spurious) scale.

The existing fixed order calculations have a natural region of validity:
when $\mu _{phy}$ is of the same order of magnitude as $m_{Q}$ and $x$ is not
too small. Our proposed scheme for heavy quark production calculation
contains the right physics when $\mu _{phy}$ becomes much larger than $m_{Q}.$
An important question to ask is then: where exactly lies the transition
region and what formalism should be used in the transition region? In order
to discuss this question in specific terms, let us choose the case of charm
production, ignoring b- and t-production completely.

Just above the threshold for producing charm, we all agree that fixed-order
calculations using flavor creation (gluon-fusion plus light-quark
scattering) alone should be reliable --- {\em provided that the parton
distribution functions used in the master formula are generated in using
evolution kernels with effective flavor number} $n_{{\it eff}}=3$! We call
this method of calculation the {\em 3-flavor scheme}. (Note, this is not
usually done in the published literature. Rather, authors of the existing
calculations invariably use canned parton distributions containing 4 or 5
quark-partons, depending on the energy scale. This procedure is, in
principle, inconsistent.) Far above threshold, say $\mu _{phy}^{2}>20\,m_{Q}^{2}$,
our proposed scheme (including charm as one of the partons) should become
the more reliable method. Here, we must use parton distributions generated
with $n_{{\it eff}}=4$ QCD evolution equation. We call this the {\em %
4-flavor scheme}. The question is: what scheme should be used in the
transition region, say $10\,m_{Q}^{2}>\mu _{phy}^{2}>20\,m_{Q}^{2}$? Although this is
one of those elusive questions in pQCD which defies definitive answer, our
best answer is: the region of transition should be considered as the region
of co-existence of the two scheme --- both schemes should be close to the
real answer and either one can provide a reasonable result. For this view
point to be viable, the answers obtained from the two approaches must be
close to each other in the transition region, with the difference being of
the order of the next order of perturbation theory, without large
logarithms. The example given in the previous section indicates that indeed
this is the case for some range of value of $\mu _{phy}$. In fact, in the
absence of reliable prediction of where the transition region lies (as with
``when should Bjorken scaling sets in?''), the requirement of approximate
equality of the prediction of the 3-flavor and 4-flavor scheme gives the
most reasonable criterion for identifying where the transition takes place.
(In the above discussion, we used $10\,m_{Q}^{2}>\mu _{phy}^{2}>20\,m_{Q}^{2}$ only as
an illustration. The appropriate numbers should be found in this
phenomenological way.)

The present paper focuses on the motivation and the physics ideas. A
detailed comparison of results from the two schemes and a study of how does
the transition takes place, as well as phenomenological results pertaining
heavy quark production at fixed-target and collider experiments will be
presented in subsequent publications.



\LaTeXparent{}

\section*{Acknowledgement}

The authors would like to thank Andrew Bazarko, Raymond Brock, Sanjib
Mishra, Jorge Morfin, Michael Shaevitz, Jack Smith, and Davison Soper for
useful discussions.

This work is partially supported by the National Science Foundation under
Grant No.PHY89-05161, the U.S. Department of Energy Contract No.
DE-FG06-85ER-40224, DE-FG05-92ER-40722, and by the Texas National Research
Laboratory Commission. M.A. and F.O. also thank the Lightner-Sams Foundation
for support. F.O. is supported in part by an SSC Fellowship.





\end{document}